\documentclass[aps,prb,twocolumn,noshowpacs]{revtex4}
\usepackage{amsmath,graphicx,amsbsy,amssymb,amsmath,epsfig,latexsym,wasysym,pifont,mathrsfs,natbib}
\usepackage{float} 
\newfloat{widefig}{thp}{lop}
\usepackage{comment} 
\usepackage{verbatim}
\usepackage{multirow,array}
\usepackage{hyperref}
\usepackage{color}
\usepackage{soul,xcolor}

\begin{document}

\newcommand{\sgcomment}[1]{\textbf{\color{green}{#1}}}
\setstcolor{red}

\title{Effect of Fe and Co substitution on martensitic stability, elastic, electronic  and magnetic properties of Mn$_2$NiGa: insights from \textit{ab initio} calculations}

\author{Ashis Kundu}
\email[]{k.ashis@iitg.ernet.in}
\author{Sheuly Ghosh}
\author{Subhradip Ghosh}
\email{subhra@iitg.ernet.in}

\affiliation{Department of Physics, Indian Institute of Technology Guwahati, Guwahati-781039, Assam, India.} 

\date{\today}

\begin{abstract}
We investigate the effects of Fe and Co substitutions on the phase stability of the martensitic phase, mechanical, electronic and magnetic properties of magnetic shape memory system Mn$_{2}$NiGa by first-principles Density functional theory(DFT) calculations. The evolution of these aspects upon substitution of Fe and Co at different crystallographic sites are investigated by computing the electronic structure, mechanical properties (tetragonal shear constant, Pugh ratio and Cauchy pressure) and magnetic exchange parameters. We find that the martensitic phase of Mn$_{2}$NiGa gradually de-stabilises with increase in concentration of Fe/Co due to the weakening of the minority spin hybridisation of Ni and Mn atoms occupying crystallographically equivalent sites. The interplay between relative structural stability and the compositional changes are understood from the variations in the elastic modulii and electronic structures. We find that like in the Ni$_{2}$MnGa-based systems, the elastic shear modulus C$^{\prime}$ can be considered as a predictor of composition dependence of martensitic transformation temperature T$_{m}$ in substituted Mn$_{2}$NiGa, thus singling it out as the universally acceptable predictor for martensitic transformation in Ni-Mn-Ga compounds over a wide composition range. The magnetic properties of Mn$_{2}$NiGa are found to be greatly improved by the substitutions due to stronger ferromagnetic interactions in the compounds. The gradually weaker(stronger) Jahn-Teller distortion (covalent bonding) in the minority spin densities of states due to substitutions lead to a half-metallic like gap in these compounds resulting in materials with high spin-polarisation when the substitutions are complete. The substitutions at the Ga site result in two new compounds Mn$_{2}$NiFe and Mn$_{2}$NiCo with very high magnetic moments and Curie temperatures. Thus, our work indicates that although the substitutions de-stabilise the martensitic phase in Mn$_{2}$NiGa, new magnetic materials with very good magnetic parameters and potentially useful for novel magnetic applications can be obtained. This can trigger interests in the experimental community for further research on substituted Mn$_{2}$NiGa.

\end{abstract}

\maketitle

%========================= introduction ==================================
\section{INTRODUCTION}

Magnetic shape memory alloys (MSMA) have drawn much attention in recent years due to their multiple functional properties such as magnetic field induced strain (MFIS), large magneto-caloric effect and magneto-resistance.~\cite{UllakkoAPL96,MurrayAPL00,SozinovAPL02,ChmielusNM09,MarcosPRB02,HuPRB01,KrenkePRB07,PasqualePRB05,BiswasAPL05,IngaleJAP09}
The MFIS is useful for magneto-mechanical actuator,\cite{UllakkoAPL96,SozinovAPL02} and large magneto-caloric effect is associated with magneto-structural coupling,\cite{AntoniJPCM09} useful for magnetic refrigeration.

Among many MSMAs Heusler Ni-Mn-Ga system has been explored extensively. The reason was that several modulated martensite phases were observed in this system with the composition ratio of Ni, Mn and Ga near 2:1:1 that is close to that of Ni$_{2}$MnGa. These modulated phases were intermediate phases during the martensitic transformation from the high temperature Heusler phase to a non-modulated tetragonal phase and had yielded MFIS as large as 6\% and 10\%.~\cite{MurrayAPL00,SozinovAPL02,LikhachevMSE04,SozinovIEEE02} However a martensitic transformation temperature (T$_{m}$) of about 200k and a Curie temperature (T$_{c}$) of about 380K\cite{WebsterPMB84} were serious hindrances for exploiting the multi-functionalities of Ni$_{2}$MnGa. This is because of the following facts: first, T$_{m}$ being lower than the room temperature makes the commercial realisation of the material for shape memory applications difficult and second, the large difference in T$_{m}$ and T$_{c}$ makes it impossible to get the maximum of the magneto-caloric effect as it requires the transition point for martensitic and magnetic transitions very close.\cite{ParetiEPJB03}

Controlled substitution of one element with another is a standard procedure for achieving target properties of materials. In order to optimise the thermodynamic parameters without compromising much on the other important properties for functional applications such as MFIS, substitution of each one of the constituents in Ni$_{2}$MnGa with other transition metals Fe, Co and Cu have been attempted. The outcome of these attempts are mixed and provide useful insights into the fundamental physics of this system which can be useful in designing this material with target properties. A larger MFIS of about 12\% was observed in the non-modulated tetragonal phase of Ni$_{2}$MnGa upon substitution of all three elements by 4\% of Co and Cu each.\cite{SozinovAPL13} Several investigations with substitution of single type of transition metal into different sub-lattices have also been carried out. The outcome of substitution at the Ni site is substantial reduction of T$_{m}$ with slight improvement in T$_{c}$,\cite{SotoPRB08,KanomataPRB09,KanomataIJAEM05} irrespective of whether the substituting element is Fe, Co or Cu. Substitution at Mn site, on the other hand, produced results depending upon the substituting element. While substitution of Mn by Co or Cu elevates T$_{m}$ and reduces T$_{c}$ as the concentration of the substituting element increases,\cite{GomesJAP06,KhanJAP05,KanomataMetals13} T$_{m}$ is observed to decrease as a function of Fe concentration when Fe is substituted at Mn site. Substitution at the Ga site by either Fe, Co or Cu shows a trend of rapid increase in T$_{m}$ and slow decrease of T$_{c}$, resulting in them coinciding for the concentration of the substituting element in the range of 10-20\%.\cite{KanomataMetals13,SotoPM10,KanomataPRB12} These outcomes, thus, prove that the structure-property relationships in Ni-Mn-Ga based system delicately depend on the substituting element, the substituent and the composition.

Mn$_{2}$NiGa is a new MSMA with the thermodynamic parameters T$_{m}$ and T$_{c}$ much better that the Ni$_{2}$MnGa, from the point of view of applications. It shows a martensitic transformation from high temperature Hg$_{2}$CuTi (Inverse Heusler) phase to a non-modulated (NM) tetragonal phase with T$_{m}$ equal to 270K,\cite{LiuAPL05,LiuPRB06} very close to room temperature. The Curie temperature, T$_{c}$, of this material is 588K,\cite{LiuAPL05,LiuPRB06} much higher than the Ni$_{2}$MnGa, which guarantees a magnetically ordered phase well above the room temperature. Both are desirable for the actuator applications at room temperature. Experimentally a MFIS of 4\% was observed in the NM tetragonal structure which was still unsaturated in a magnetic field of 1.8T\cite{LiuAPL05} implying that a larger MFIS can be obtained with larger filed. In a recent Density functional theory (DFT) based investigations, a number of modulated phases were predicted,\cite{Kundumodulation17} some of which were observed in experiments\cite{SinghAPL10,BrownJPCM10} as well. The DFT calculations predicted that larger MFIS can be realized in the modulated phases. Very recently, a large inverse magnetocaloric effect is also reported in this system.\cite{SinghAPL14} Therefore, Mn$_{2}$NiGa qualifies the requirements of a MSMA with multiple functionalities, often better than the prototype Ni$_{2}$MnGa.

Inspite of having reasonable and more desirable functional properties, one crucial issue with Mn$_{2}$NiGa is it's low value of magnetisation which is about 1.2$\mu_{B}$ per formula unit as opposed to more than 4$\mu_{B}$ per formula unit in Ni$_{2}$MnGa. This happens due to a ferrimagnetic ground state arising out of anti-parallel orientation of the two Mn atoms. Substitution by another magnetic atom like Co, Fe and Cu could be an useful way to adjust the magnetic interactions in the parent alloys, thus improving the magnetisation primarily. With this motivation, Luo \textit{et al.} substituted Mn with Fe in Mn$_{2}$NiGa.\cite{LuoJAP10} They observed an increase in the saturation magnetisation with increasing Fe concentration. However, both T$_{m}$ and T$_{c}$ decrease with increasing Fe content and no martensitic transformation is observed beyond 30\% of Fe content\cite{LuoJAP10}. Ma \textit{et al.} investigated the effect of Co substitution at Ni and Ga sites of Mn$_{2}$NiGa through magnetisation measurement and \textit{ab initio} calculations.\cite{MaPRB11} They found a remarkable three times jump in the saturation magnetisation when Co is substituted at Ga sites which, by means of \textit{ab initio} calculation, was attributed to a complex sub-lattice occupancy pattern. Though Co substitution at Ni site wasn't as dramatic, the magnetisation improved which was explained by means of increasing ferromagnetic component in a ferrimagnetic host. In both cases, the martensitic transformation vanished rapidly indicating stabilisation of the the Inverse Heusler phase down to low temperature. A different variation of T$_{c}$ with Co content was observed depending on the site of substitution. Very recently, DFT calculations on Cu doped Mn$_{2}$NiGa reported that T$_{m}$ decreases when Cu is substituted at Mn and Ni sites but increases when substituted at Ga site.

The investigations on substitution of another transition metal in Mn$_{2}$NiGa are thus quite scattered. However, they offer very interesting perspectives, both for fundamental understanding as well as for engineering materials with target properties. The number of valence electron per atom(e/a) was identified to be a predictor of T$_{m}$ with T$_{m}$ ${\sim}$ e/a\cite{DoCalphad08,RamamurthyJPCS86,WinderJPCS58} for systems undergoing martensitic transformations. In case of off-stoichiometric Ni$_{2}$MnGa alloys and Fe, Co and Cu substituted Ni$_{2}$MnGa alloys e/a was found not to correlate with T$_{m}$.\cite{HuPRB09,ChunmeiPRB10,ChunmeiPRB11} Instead, the shear modulus C$^{\prime}$ was found to be a better predictor for composition dependence of T$_{m}$.\cite{ChunmeiPRB10,ChunmeiPRB11} On the other hand, $\Delta$E, the energy difference between the high temperature Heusler phase and the low temperature NM tetragonal phase, was found to correlate well with C$^{\prime}$ and T$_{m}$ for a number of systems in the Ni-Mn-Ga family.\cite{GhoshPBCM11,BarmanPRB08Ga2} Experimental results on substituted Mn$_{2}$NiGa\cite{LuoJAP10,MaPRB11} indicate that e/a does not correlate with T$_{m}$. On the other hand, the variations of T$_{c}$ and magnetisation in substituted Mn$_{2}$NiGa depend, both quantitatively as well as qualitatively, not only on the nature of the element that is being substituted. Another noteworthy point is the gradual stabilisation of the high symmetry Inverse Heusler phase with substitution, irrespective of the chemical identity of the substituting atom and the site in which it is being substituted, in substituted Mn$_{2}$NiGa. One, therefore, needs a systematic first-principle based investigation addressing the multiple issues, under a single approximation, in order to provide fundamental understanding of the interrelations between composition, sub-lattice occupancy, phase stability and magnetic interactions. An investigation along this line would help to tune the necessary parameters for targeted applications in Mn$_{2}$NiGa base systems and possibly in Ni-Mn-Ga systems over a wide composition range.

In this paper, we report the outcome of substitution of 25\%, 50\%, 75\% and 100\% Fe and Co at different sites of the parent compound Mn$_{2}$NiGa. Specifically we have looked at the patterns of site occupancies upon a particular substitution, the stabilities of the martensitic NM phases, the elastic properties, the total and atomic magnetic moments, the effective magnetic exchange interactions and the magnetic transition temperatures and their variations upon substitution at different sites. The results are interpreted from the composition dependencies of the computed electronic structures. This approach enables us to pinpoint the microscopic origin of the martensitis phase stability upon different substitutions, the variations in the magnetic properties with compositions and nature of the substitutions, the variations in the mechanical properties and their interrelations with the nature of the martensitic stability and most importantly in establishing a predictor for variations in T$_{m}$. Another outcome of this investigation is the prediction of a new material with high magnetisation and high T$_{c}$ which can be used in various applications which require large moment and stability of magnetically ordered phase and which can fall in the same class of X$_{2}$YZ materials with Heusler or Heusler-like structures where all three components have unfilled d-shells.\cite{SanvitoSA17}

The paper is arranged as follows: In Sec. II, the first principles computational method and the calculational details are provided. In Sec. III, the results of our calculations and their analysis are presented. The final remarks on this work are presented in Sec. IV.

\section{Computational Methods}
The total energies, the electronic structures and the magnetic moments were calculated  with spin-polarized
density functional theory (DFT) based projector augmented wave (PAW)
method as implemented in Vienna Ab-initio Simulation Package
(VASP).\cite{PAW94,VASP196,VASP299} For all calculations, we have used
Perdew-Burke-Ernzerhof implementation of Generalised Gradient
Approximation (GGA) for exchange-correlation functional.\cite{PBEGGA96} An
energy cut-off of 450 eV and a Monkhorst-Pack\cite{MP89} $11 \times 11 \times11$
$k$-mesh were used for self consistent
calculations. A larger $k$-mesh of $15 \times 15 \times15$ was used for the calculations of the electronic structures. The convergence criteria for the total energies and the forces on individual atoms were set to 10$^{-6}$ eV and $10^{-2}$ eV/\r{A} respectively. To
investigate the stability of the substituted compounds, we have calculated the
formation energy for each system, which can be obtained the following way,

\begin{eqnarray}
E_{f}=E_{tot}-\sum_{i}n_{i}E_{i}
\end{eqnarray}
 $E_{tot}$ is the ground state total energy of a system, $E_{i}$ is the ground state energy of the $i$-th
 component in it's elemental phase and $n_{i}$ is it's concentration in the system under consideration.
The elastic constants for the compounds are calculated only for their high temperature phases with cubic symmetry. The details of the calculations are given in the supplementary material.

The magnetic pair exchange parameters are computed in order to understand the nature of the magnetic interactions in these systems.  They are efficiently calculated using multiple
scattering Green function formalism as implemented in SPRKKR
code.\cite{EbertRPP11} In this approach, the spin part of the Hamiltonian is
mapped to a Heisenberg model

\begin{eqnarray}
H = -\sum_{\mu,\nu}\sum_{i,j}
J^{\mu\nu}_{ij}
\mathbf{e}^{\mu}_{i}
.\mathbf{e}^{\nu}_{j}
\end{eqnarray}
$\mu$, $\nu$ represent different sub-lattices, \emph{i}, \emph{j}
represent atomic positions and $\mathbf{e}^{\mu}_{i}$ denotes the unit
vector along the direction of magnetic moments at site \emph{i}
belonging to sub-lattice $\mu$. The $J^{\mu \nu}_{ij}$s are calculated
from the energy differences due to infinitesimally small orientations
of a pair of spins within the formulation of Liechtenstein \textit{et al.}.\cite{LiechtensteinJMMM87} In order to calculate the energy
differences by the SPRKKR code, full potential spin polarised scaler
relativistic Hamiltonian with angular momentum cut-off $l_{max} = 3$
is used along with a converged $k$-mesh for Brillouin zone
integrations. The Green's functions are calculated for 32 complex
energy points distributed on a semi-circular contour. The energy
convergence criterion was set to 10$^{-5}$ eV for the self-consistency
cycles. The equilibrium lattice parameters  obtained from the PAW calculations were used in these calculations. These exchange parameters are then used for the calculations of Curie temperatures (T$_{c}$). The Curie temperatures are estimated with two different approaches: the Mean field approximation (MFA)\cite{SokolovskiyPRB12} and the Monte Carlo simulation (MCS) method in order to check the qualitative consistency in the results and to obtain a reliable estimate of the
quantity as the MFA is known to overestimate T$_{c}$ while the MCS method is more accurate quantitatively. Details of the MFA and MCS calculations are given in the supplementary material.

%====================Results and Discussions===============================

\section{Results and Discussions}

At high temperature, Mn$_{2}$NiGa crystallises in Hg$_{2}$CuTi (Inverse Heusler) structure (space group no. 216; $F\bar{4}3m$) with four inequivalent Wyckoff positions (4a, 4b, 4c, 4d) in an FCC unit cell.~\cite{LiuAPL05,PaulJAP11} The Mn atoms occupy the 4a (0,0,0) and 4c (0.25, 0.25, 0.25) Wyckoff positions; we denote them as MnI and MnII respectively. The 4b (0.5, 0.5, 0.5) and 4d (0.75, 0.75, 0.75) positions are occupied by Ni and Ga respectively. In this work, we focus on the high temperature phase as results obtained in this phase would be enough for most of the physical understanding about the effects of substituting another magnetic element on the functional properties of Mn$_{2}$NiGa. To model the chemical substitution, we have taken a 16 atom conventional cubic cell. Thus, chemical substitution of 25 \%, 50\%, 75\% and 100\% can be modelled by successive replacement of the atoms of one of the constituents. For example, to make a $25\%$ of Co substitution at Ni site, one Ni atom out of the four in the 16 atom cell is to be replaced with one Co atom. This modelling strategy has worked well in cases of investigations on chemically substituted Ni$_{2}$MnGa.\cite{BuchelnikovJPDAP15,SokolovskiyPRB15}

\subsection{Site preferences, stability and structural parameters}
\begin{figure}[t]
\centerline{\hfill
\psfig{file=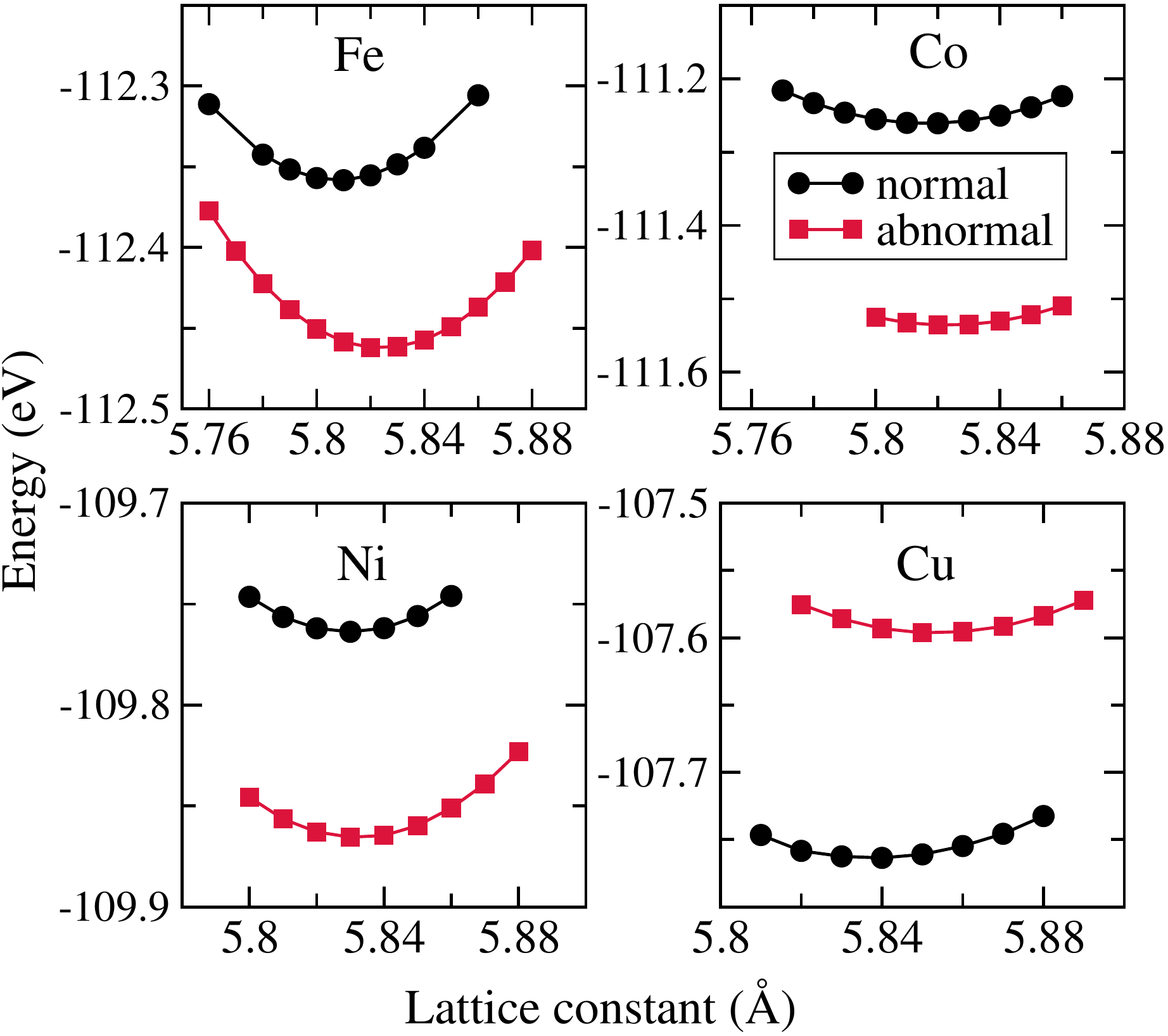,width=0.43\textwidth}\hfill}
\caption{Calculated total energy as a function of lattice constant for 25\% Fe, Co, Ni and Cu substituted at Ga site in Mn$_{2}$NiGa(Mn$_{2}$NiGa$_{0.75}$X$_{0.25}$). ``normal'': Fe, Co, Ni or Cu occupy the Ga sublattice; ``abnormal'': Fe, Co, Ni or Cu occupy the MnI sites and remaining MnI atoms move to Ga sites.}
\label{latc}
\end{figure}

The site occupancies in a substituted system have important impact on the physical properties of the system. Thus, before proceeding with computations of the physical properties, the site preferences of the substituting atom is to be decided. If the substituting atom occupies the site of the substituted element, the configuration is ``normal'', else it is termed ``abnormal''. The preferred site occupancy is determined by comparing the total energies of the two configurations.
 In case of Fe and Co-substituted Ni$_{2}$MnGa, free energy calculations revealed that the preferred configuration in cases of substitutions at Ni or Mn sites are ``normal'' whereas the substitution at Ga site prefers an ``abnormal'' configuration(Substituting Fe(Co) occupies the Mn(Ni)sites while the rest of  Mn(Ni) move to Ga sites).\cite{ChunmeiPRB11} For Cu substituted Ni$_{2}$MnGa, the preferred configuration always is the ``normal'' one.\cite{ChunmeiPRB11} In case of Co and Cu substituted Mn$_{2}$NiGa, the trend of site occupancy is found to be quite similar to that of substituted Ni$_{2}$MnGa systems.\cite{MaPRB11,ChakrabartiPRB13} It has been observed that in Heusler alloys, the following pattern of site occupancy is generally followed: the substituting transition metal atom will prefer the 4a and 4b sites if it has a larger number of valence electrons, while it will prefer the 4c and 4d sites if the number of valence electrons is less.\cite{LiuPRB08,KandpalJPD07,FengPRB01,BurchPRL74,HelmholdtJLCM87} This has been observed even in cases of anti-site disorder in Mn$_{2}$NiZ alloys.\cite{PaulJPCM13,PaulJAP14} So the substituting Fe or Co in Mn$_{2}$NiGa would prefer the MnI sites over MnII sites if Mn atoms are being substituted, as, Fe and Co both have larger number of valence electrons than Mn. We have  verified this by comparing the total energies of two cases: one, where the entire Fe/Co occupies MnI sites and two, Fe/Co are equally distributed among the two Mn sites. For substitution of Fe or Co at Ni site, we find the ``normal'' site occupancy(Fe/Co occupying Ni sites) to be  energetically favourable. This is consistent with the general pattern described above and the previous first-principles results on Co substituted Mn$_{2}$NiGa.\cite{MaPRB11} The substitution at Ga site, however, follows a different pattern, depending upon the substituting element. In case of Co substitution at Ga site in Mn$_{2}$NiGa, earlier work \cite{MaPRB11} showed that the Co prefers to occupy the MnI sites pushing the remaining MnI atoms to Ga sites(henceforth denoted as MnIII). This can be understood on the basis of the general occupancy pattern in Heusler alloys described above:  since Co has more valence electrons than Mn and Ga, it will occupy the MnI sites and the remaining MnI will occupy the Ga sites and would be distinguished from MnI and MnII by being denoted as MnIII. In order to check whether this is indeed the cases with both Fe and Co-substituted system, we have computed the total energies for ``normal'' and ``abnormal'' configurations. The results for $25\%$ substitution are shown in Fig. \ref{latc}. For comparison, we have also shown the results for Ni and Cu-substitution at Ga sites. The results suggest that the ``abnormal'' site occupancies are preferable for Fe, Co and Ni substitutions at Ga sites of Mn$_{2}$NiGa while Cu substitution prefers a ``normal'' configuration. This exactly follows the trend obtained in substituted Ni$_{2}$MnGa \cite{ChunmeiPRB11}.  We can thus conclude that the site preferences of the substituting transition metal atom in Ni-Mn-Ga systems is dependent on the valence shell electronic configurations of both the atom that is being substituted and the substituting atom.

\begin{figure}[t]
\centerline{\hfill
\psfig{file=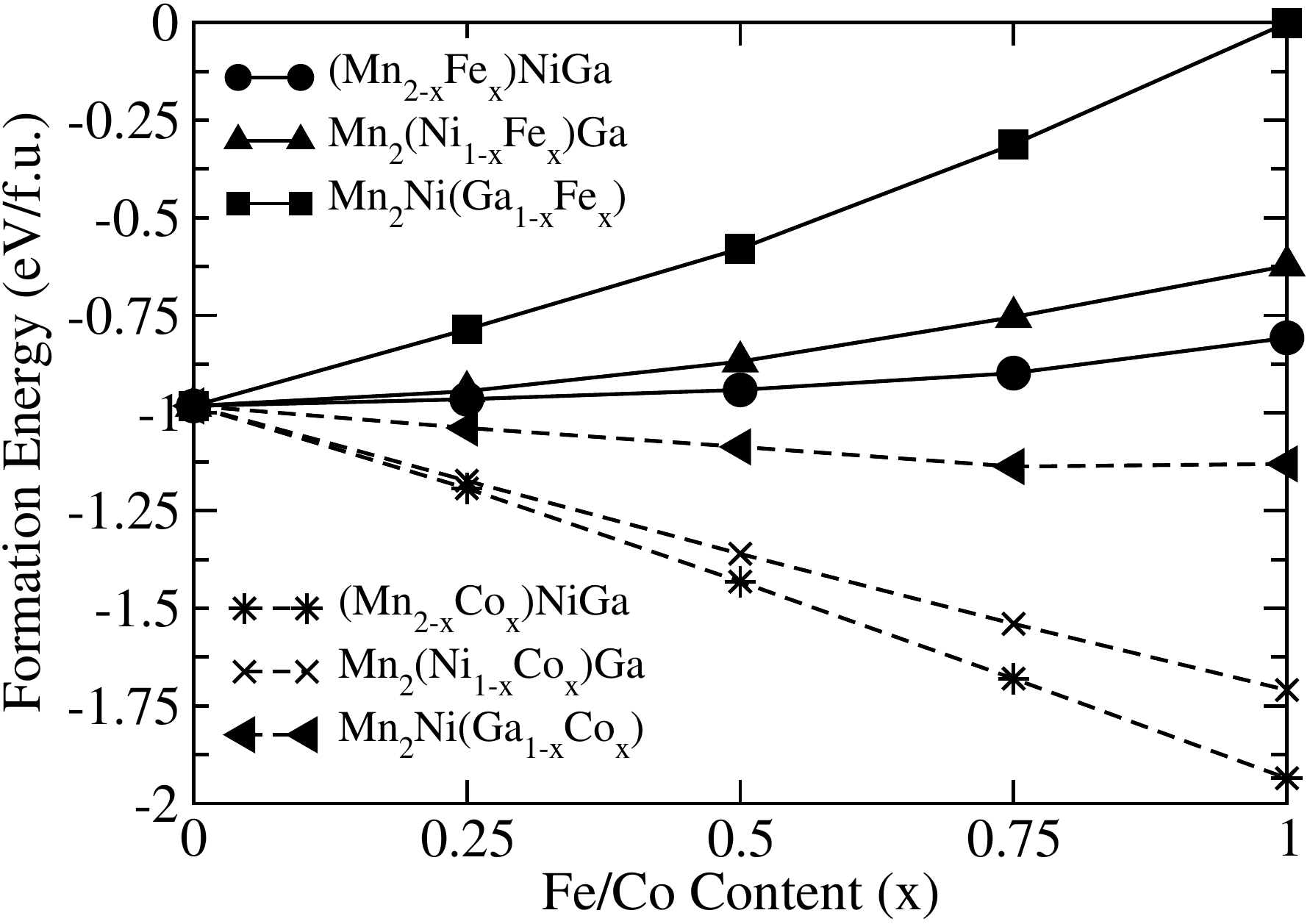,width=0.4\textwidth}\hfill}
\caption{Formation energy (eV/f.u.) as a function of Fe and Co content
 substituted at different sites in Mn$_{2}$NiGa.}
\label{form}
\end{figure}

\begin{table*}[t]
\centering

\caption{\label{table1} Calculated values of equilibrium lattice constants (a$_{0}$), electron to atom ratios ($e/a$) and the formation energies (E$_{f}$) of various compounds in the Hg$_{2}$CuTi phase are given. The total energy differences ($\Delta{E}$) between the austenite (Hg$_{2}$CuTi) phase and the martensite (tetragonal) phase (The equilibrium value of $(c/a)$, the tetragonal distortion, is given in parentheses) and the corresponding volume changes ($|\Delta{V}|/V$) with respect to the equilibrium volume in the Hg$_{2}$CuTi phase, are given in 5th and 6th column. M$_{\rm A}$ is the total magnetic moment in $\mu_{B}$/f.u. of the austenite phase. Reported values of lattice constants and magnetic moments in the literature  are also given.}
\resizebox{1.0\textwidth}{!}{%
\begin{tabular}{l@{\hspace{0.4cm}} c@{\hspace{0.4cm}} c@{\hspace{0.4cm}} c@{\hspace{0.4cm}} c@{\hspace{0.4cm}} c@{\hspace{0.4cm}} c@{\hspace{0.4cm}} c@{\hspace{0.4cm}} c@{\hspace{0.4cm}} }
\hline\hline
\vspace{-0.33 cm}
\\ System  & a$_{0}$(\r{A}) & e/a &  E$_{\rm f}$ &  $\Delta{E}$(c/a) &  $|\Delta{V}|/V$  & M$_{\rm A}$ & a$^{\rm Lit.}_{0}$(\r{A}) & M$^{Lit.}_{\rm A}$  \\
  & & &  (eV/f.u.) &  (mev/atom) &  (\%)  & ($\mu_{\rm B}/f.u.$) &  & ($\mu_{\rm B}/f.u.$)   \\ \hline\hline

  Mn$_{2}$NiGa                     & 5.84 & 6.75 & -0.98 &  26.98(1.28) & 0.65 &  1.16 & 5.91,\cite{LiuAPL05,LuoJAP10} 5.85,\cite{BarmanEPL07,WollmannPRB14} 5.88\cite{MaPRB11}  & 1.14,\cite{BarmanEPL07} 1.18\cite{MaPRB11} \\
  (Mn$_{0.75}$Fe$_{0.25}$)NiMnGa   & 5.83 & 6.8125 & -0.96 & 8.55(1.26) & 0.98 &  1.72 &  5.88\cite{LuoJAP10} & 1.55\cite{LuoJAP10}   \\
  (Mn$_{0.5}$Fe$_{0.5}$)NiMnGa    & 5.81 & 6.875 & -0.94 & - & - &  2.62 &  5.86\cite{LuoJAP10} &   2.68\cite{LuoJAP10} \\
  (Mn$_{0.25}$Fe$_{0.75}$)NiMnGa   & 5.78 & 6.9375 & -0.90 & - & - &  3.46 &   &   \\
  FeNiMnGa                         & 5.75 & 7 & -0.81 & - & - &  4.02 & 5.799\cite{AlijaniPRB11}  &  4.01\cite{AlijaniPRB11} \\
\hline
  Mn$_{2}$NiGa                         & 5.84 & 6.75  & -0.98 & 26.98(1.28) & 1.04 & 1.16  &   &   \\
  Mn$_{2}$(Ni$_{0.75}$Fe$_{0.25}$)Ga   & 5.82 & 6.625 & -0.94 & 11.80(1.30) & 0.42 & 1.45  &   &   \\
  Mn$_{2}$(Ni$_{0.5}$Fe$_{0.5}$)Ga     & 5.80 & 6.5   & -0.87 & 5.07(1.34)  & 1.67 & 1.49  &   &    \\
  Mn$_{2}$(Ni$_{0.25}$Fe$_{0.75}$)Ga   & 5.79 & 6.375 & -0.75 & 15.04(1.38) & 1.71 & 1.31  &   &    \\
  Mn$_{2}$FeGa                         & 5.78 & 6.25  & -0.62 & 33.07(1.40) & 2.53 & 1.04  & 5.80,\cite{LuoJAP08}  5.76\cite{WollmannPRB14}  &  1.03\cite{WollmannPRB14}  \\
\hline
 Mn$_{2}$NiGa                          & 5.84 & 6.75   & -0.98 & 26.98(1.28) & 0.65 & 1.16  &   &  \\
  Mn$_{2}$Ni(Ga$_{0.75}$Fe$_{0.25}$)   & 5.82 & 7.0625 & -0.78 & 9.56(1.30) & 1.82 & 2.79  &   &  \\
  Mn$_{2}$Ni(Ga$_{0.5}$Fe$_{0.5}$)     & 5.80 & 7.375  & -0.58 & - & -  &  4.63 &   &  \\
  Mn$_{2}$Ni(Ga$_{0.25}$Fe$_{0.75}$)   & 5.77 & 7.6875 & -0.31 & - & -   & 6.40  &   &  \\
  Mn$_{2}$NiFe                         & 5.74 & 8      & 0.00 & - & -  & 8.03  &   &  \\
\hline
   Mn$_{2}$NiGa                         & 5.84 & 6.75  & -0.98 & 26.98(1.28) & 0.65  & 1.16  &   &  \\
  (Mn$_{0.75}$Co$_{0.25}$)NiMnGa       & 5.83 & 6.875 & -1.19 & 18.57(1.26) & 0.43  &  1.97 &   &  \\
  (Mn$_{0.5}$Co$_{0.25}$)NiMnGa        & 5.81 & 7     & -1.43 & - & - &  2.95 &   &  \\
  (Mn$_{0.25}$Co$_{0.75}$)NiMnGa       & 5.80 & 7.125 & -1.68 & - & -  & 3.78  &   &  \\
  CoNiMnGa                             & 5.78 & 7.25  & -1.94 & - & -  &  4.98 & 5.803\cite{AlijaniPRB11}  & 5.07\cite{AlijaniPRB11} \\
\hline
  Mn$_{2}$NiGa                         & 5.84 & 6.75   & -0.98 & 26.98(1.28) & 0.65  & 1.16  &   &  \\
  Mn$_{2}$(Ni$_{0.75}$Co$_{0.25}$)Ga   & 5.82 & 6.6875 & -1.17 & 18.51(1.28) & 0.02  & 1.46  &   & 1.52\cite{MaPRB11} \\
  Mn$_{2}$(Ni$_{0.5}$Co$_{0.5}$)Ga     & 5.80 & 6.625  & -1.36 & 7.00(1.28) & 1.02   & 1.71  & 5.88\cite{MaPRB11}  & 1.70\cite{MaPRB11} \\
  Mn$_{2}$(Ni$_{0.25}$Co$_{0.75}$)Ga   & 5.78 & 6.5625 & -1.54 & - & -   & 1.92  &   &  \\
  Mn$_{2}$CoGa                         & 5.76 & 6.5    & -1.70 & - & -  &  2.00 & 5.78\cite{WollmannPRB14}  & 2.00\cite{WollmannPRB14} \\
\hline
  Mn$_{2}$NiGa                         & 5.84 & 6.75  & -0.98 & 26.98(1.28) & 0.65  & 1.16  &   &  \\
  Mn$_{2}$Ni(Ga$_{0.75}$Co$_{0.25}$)   & 5.82 & 7.125 & -1.04 & 16.99(1.30) & 1.82  & 2.88  &   & 3.11\cite{MaPRB11} \\
  Mn$_{2}$Ni(Ga$_{0.5}$Co$_{0.5}$)     & 5.81 & 7.5   & -1.09 & - & -  & 4.73  & 5.84\cite{MaPRB11}  & 5.29\cite{MaPRB11} \\
  Mn$_{2}$Ni(Ga$_{0.25}$Co$_{0.75}$)   & 5.80 & 7.875 & -1.14 & - & -   & 6.94  &   &  \\
  Mn$_{2}$NiCo                         & 5.79 & 8.25  & -1.13 & - & -   &  9.04 &   &  \\
              
\hline\hline
\end{tabular}
}
\end{table*}

\begin{figure*}[t]
\centerline{\hfill
\psfig{file=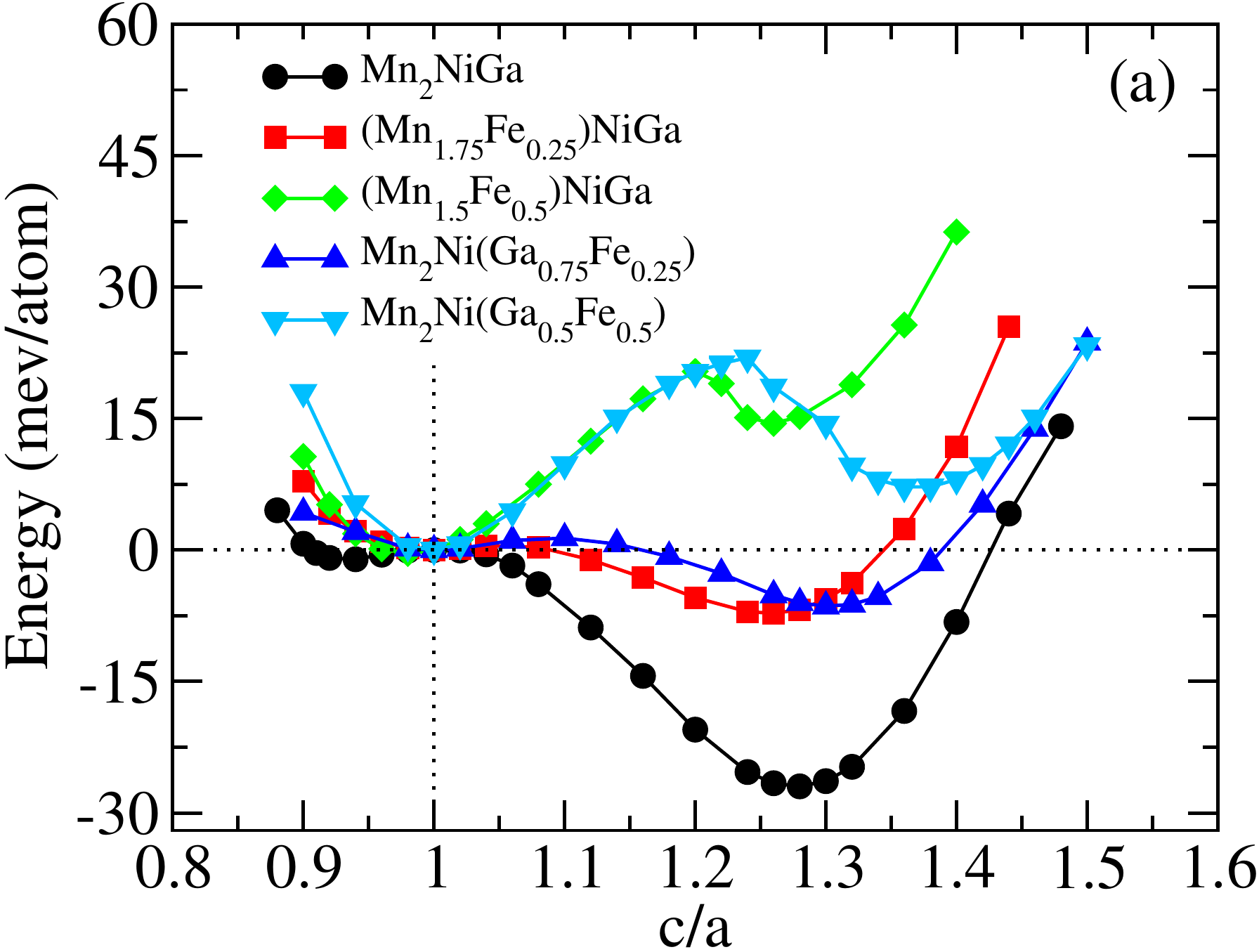,width=0.35\textwidth}
\hspace{0.5cm}
\psfig{file=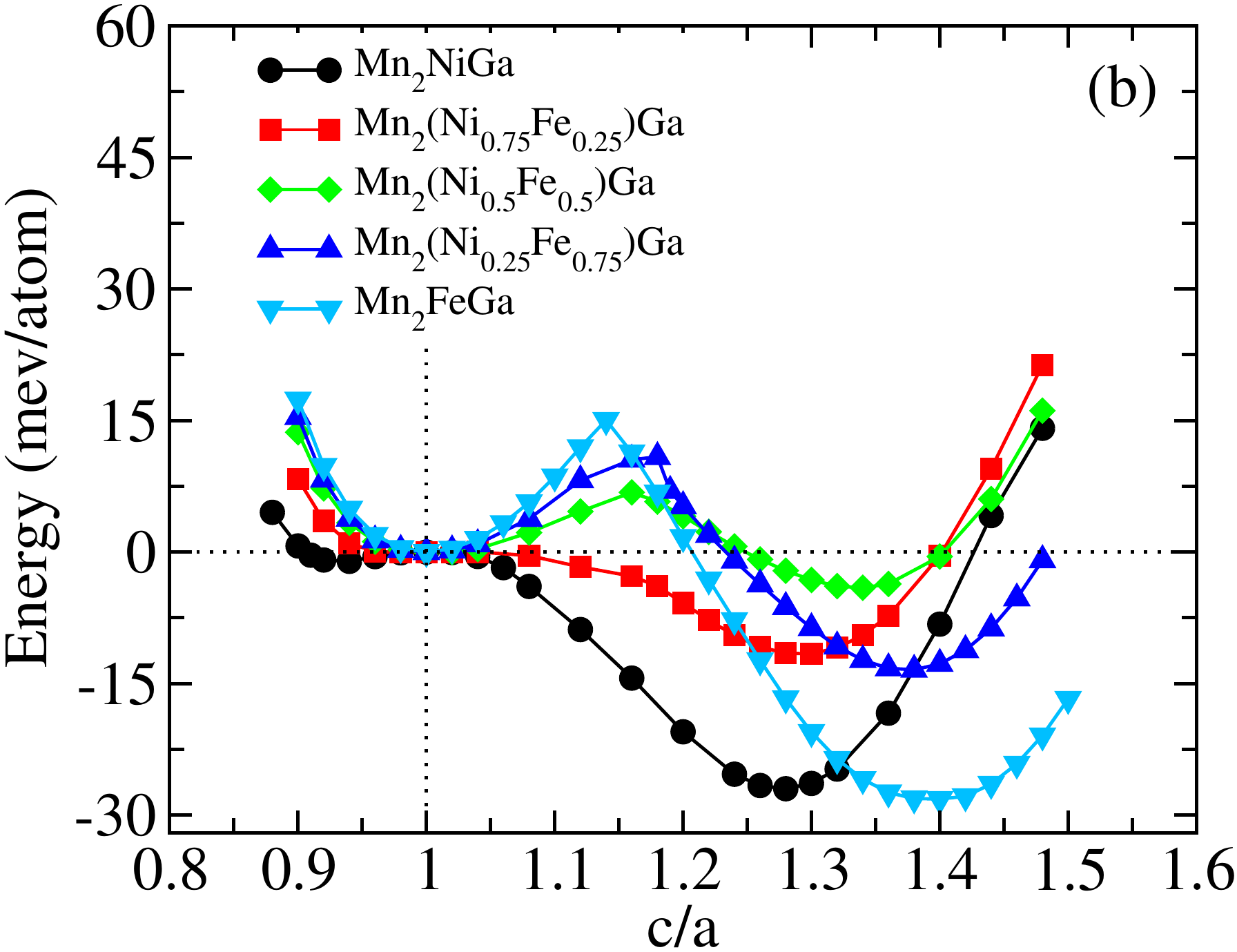,width=0.35\textwidth}\hfill}
\vspace{0.5cm}
\centerline{\hfill
\psfig{file=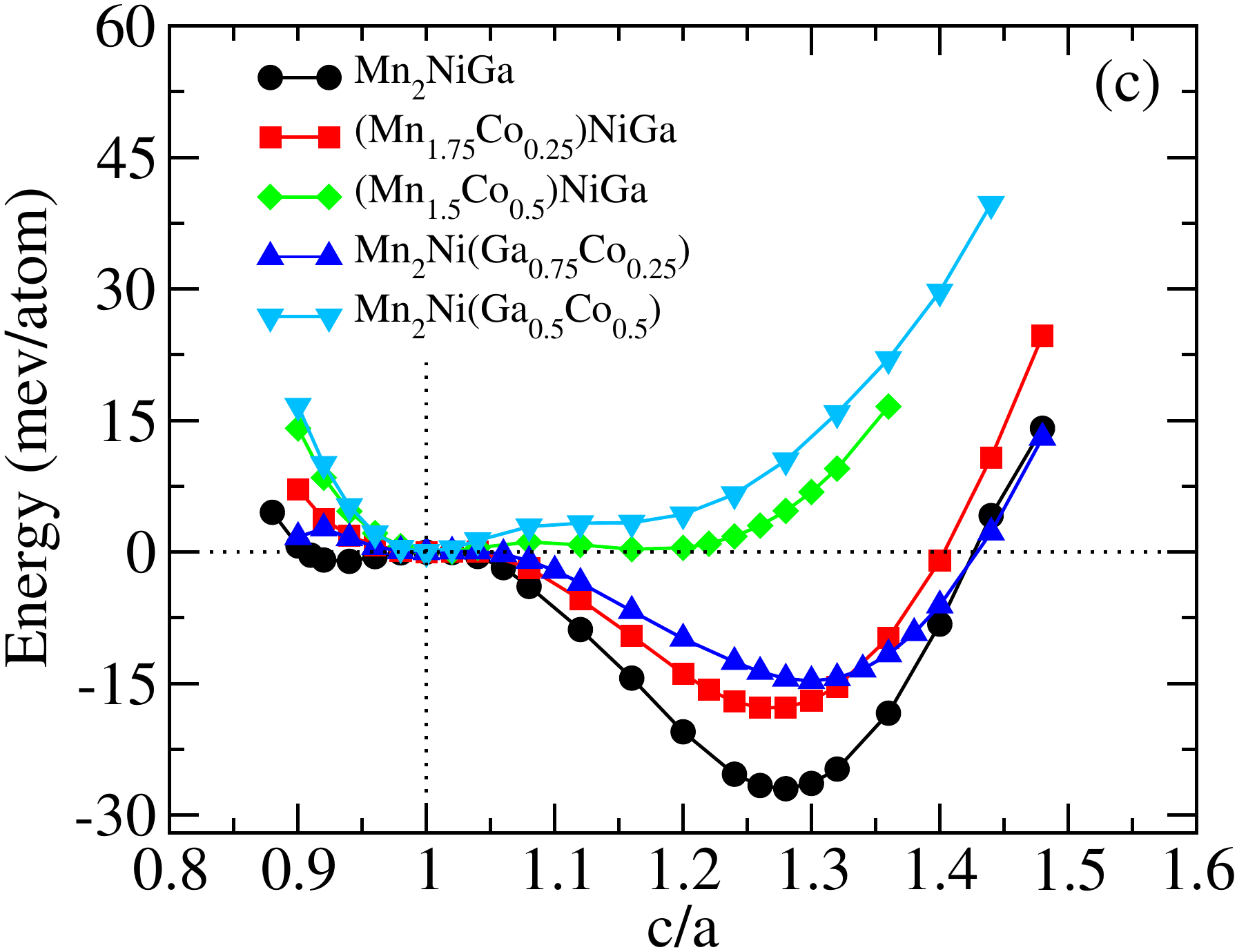,width=0.35\textwidth}
\hspace{0.5cm}
\psfig{file=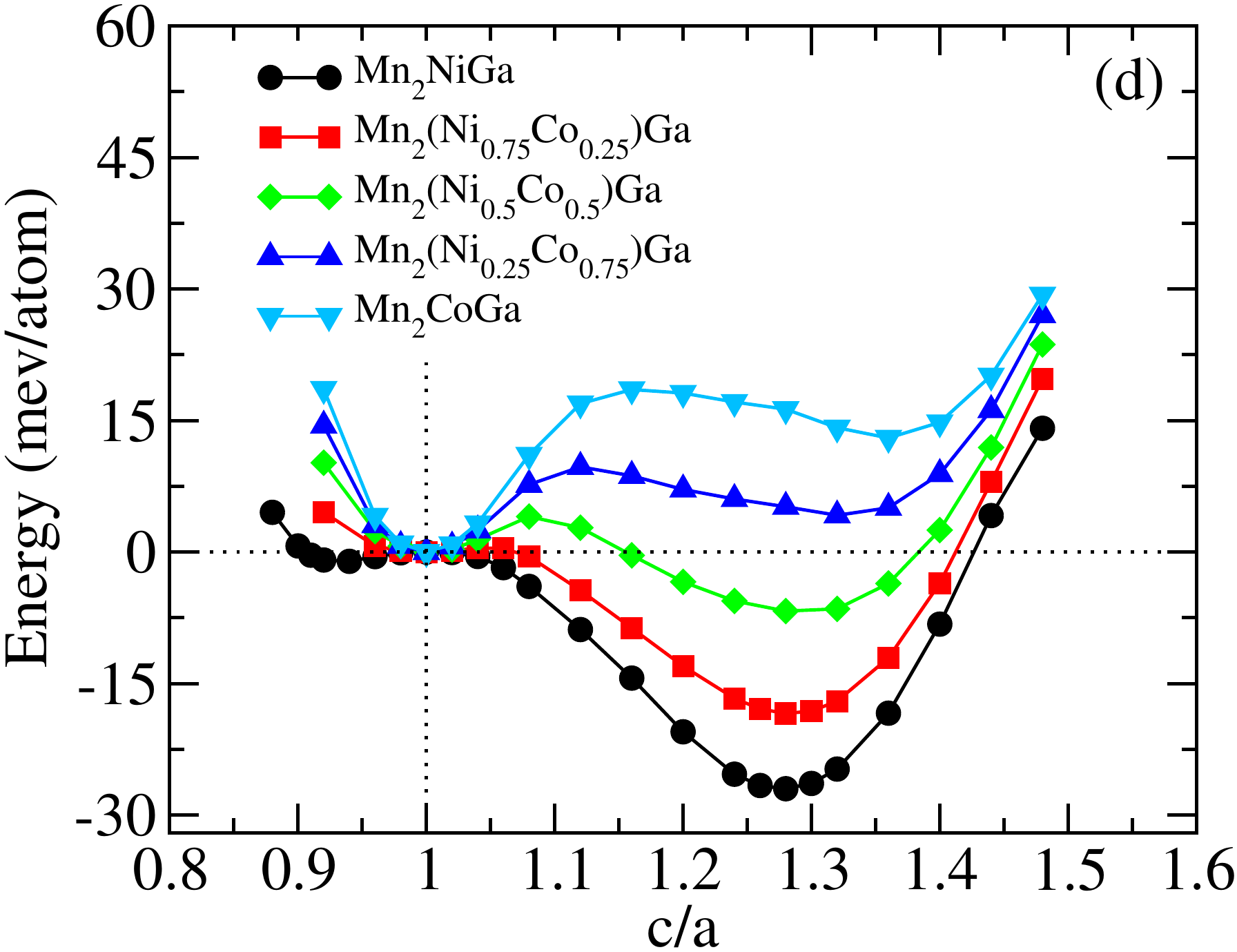,width=0.35\textwidth}\hfill}
\caption{The variations of total energies as a function of c/a ratio for Fe and
  Co substituted Mn$_{2}$NiGa compounds. The zero energy is
  taken to be the energy of austenite phase.}
\label{tot-en}
\end{figure*}

After fixing the site preferences of the atoms in substituted Mn$_{2}$NiGa, we computed the equilibrium lattice constants and the formation energies of the compounds obtained by chemical substitutions at various sites in the cubic Hg$_{2}$CuTi phase. The results are presented in Table~\ref{table1}.
For all the cases, the equilibrium lattice constants decreases linearly with increasing concentration of the substituting element. For most of the cases, this trend can be explained from the variations of atomic radii of host (The atomic radii of Mn, Ni and Ga are 1.27\r A, 1.24\r A and 1.35\r A respectively) and the substituting elements(The atomic radii of Fe and Co are 1.26\r A and 1.25\r A respectively). The only exception are the substitutions at the Ni site where instead of an expected increase in the lattice constant with concentration of the substituting element, the lattice constant decreases. Thus, the trends in the variations in the lattice constants cannot be understood in terms of differences in atomic radii alone, and the other effects like bonding and magnetism are expected to play roles as were noted earlier.\cite{ChunmeiPRB11,ChakrabartiPRB13} Another noteworthy point is that while the trends in the variations of the lattice constants obtained in this calculations qualitatively agree with that observed in the experiments on Fe-substituted at Mn sites \cite{LuoJAP10} and Co substituted at Ga sites of Mn$_{2}$NiGa, \cite{MaPRB11} the experimentally observed trends are opposite to our calculated results in cases of Co-substituted at the Ni sites. \cite{MaPRB11} Experimentally, it is found that the  lattice constant increases with increasing concentration of Co when it is substituted at the Ni site, although the increase is very slow (about $0.1\%$ maximum). Thus, the calculated trends are consistent with the experiments for most of the systems under consideration here.
 
In Fig. \ref{form}, we present the variations in  formation energies of compounds  with Fe or Co substituted at different sites in Mn$_{2}$NiGa as a function of the concentration of the substituent. It can be seen that the formation energy is negative for all systems except Mn$_{2}$NiFe(E$_{f}$=0). This implies that except Mn$_{2}$NiFe all the compounds can form in the  the Hg$_{2}$CuTi structure from  enthalpy point of view. We haven't come across any experimental result regarding Mn$_{2}$NiFe contradicting this finding. As some of the  compounds under investigation such as Mn$_{2-x}$Fe$_{x}$NiGa(x=0-0.5), \cite{LuoJAP10} Mn$_{2}$Ni$_{1-x}$Co$_{x}$Ga(x=0-0.5) and Mn$_{2}$NiGa$_{1-x}$Co$_{x}$(x=0-0.52)\cite{MaPRB11} have already been synthesised, our DFT calculations correctly reproduce the experimental observations and hence can provide a guidance regarding possibility of synthesising the ones which have not been synthesised yet. Regarding the relative stabilities of compounds upon substitution of different atoms, our calculations imply that the substitution at different sites by Co is more favourable than substitution by Fe. Also the substitutions at the Ga sites make the compounds least stable. By comparing the formation energies calculated here and in Ref. \onlinecite{ChakrabartiPRB13}, we conclude that the Co substitution in Mn$_{2}$NiGa produces more stable compounds compared to the ones formed by substituting Fe or Cu.

\subsection{Martensitic phase transformation}
From the point of view of functionality, it is important to investigate the sustainability of the martensitic transition of Mn$_{2}$NiGa upon substitutions by various atoms. Experiments  on Co-substitution at Ni and Ga sites \cite{MaPRB11} did not find any martensitic transformation beyond $16\%$ of Co in the system. The measurements on Mn$_{2}$Co$_{0.08}$Ni$_{0.92}$Ga and Mn$_{2}$Co$_{0.08}$NiGa$_{0.92}$  showed that T$_{m}$ decreases rapidly to 125K and 103K respectively from 270K in case of Mn$_{2}$NiGa. Fe-substitution at Mn site also led to a rapid decrease of T$_{m}$(120K for Mn$_{1.7}$Fe$_{0.3}$NiGa) with no trace of martensitic transformation being observed beyond $30\%$ of Fe. In this sub-section, we examine the stabilities of the martensitic phases for all the compounds in order to find if there is any trend with quantities like (e/a) so that a predictor of variation in the T$_{m}$ can be fixed. We do this by computing the total energy differences, $\Delta E$, between the high temperature Hg$_{2}$CuTi phases and the low temperature non-modulated martensitic phases with tetragonal structure, the later being obtained by tetragonal distortion of the former. While $\Delta E$ does not provide accurate quantitative estimate of T$_{m}$ as that requires inclusion of various contributions to the free energy which are difficult to compute for a chemically disordered system, it's variations with compositions help make a heuristic predictions on qualitative variations of T$_{m}$ and the stability of the martensitic phases. Such an approach has been adopted elsewhere~\cite{GhoshPBCM11,ChakrabartiPRB13} successfully.

In Fig. \ref{tot-en}, we show the results of variations in the total energies with tetragonality(c/a) by keeping the volume constant at the equilibrium volume of Hg$_{2}$CuTi phases for each one of Fe and Co substituted  Mn$_{2}$NiGa systems. The reference energy in all cases is fixed at the one corresponding to (c/a)=1, the Hg$_{2}$CuTi phase. From Fig. \ref{tot-en}(a) and (c), it can be seen that the Fe or Co substitutions at Mn or Ga sites reduce the stabilities of the martensite phases as the values of $\Delta{E}$ decrease(see Table-\ref{table1}) compared to the host Mn$_{2}$NiGa($\Delta{E}$=26.98 meV/atom) as the concentrations of the substituents increase. The total energy plots for (Mn$_{2-x}$Fe$_{x}$)NiGa show that for $x=0.25$, the martensitic phase is almost de-stabilised; the $\Delta E$ decreases by a factor of more than 3 in comparison to the parent compound. At $x=0.5$, the tetragonal phase is not even energetically favourable. These are in good agreement with the experimental observations as $\Delta E$ is considered as the predictor for T$_{m}$.\cite{LuoJAP10} In case of Mn$_{2}$Ni(Ga$_{1-x}$Co$_{x}$), our calculated trends on the composition dependence of the martensitic transformation differs slightly from the experimental observations. In the experiments, no martensitic transformation was observed beyond $x=0.16$. \cite{MaPRB11} Our calculations, on the other hand, obtains a substantially deep energy minima at (c/a) $\neq$ 1 for $x=0.25$, signifying the possibility of martensitic transformation at this composition. Qualitatively, though, this result is in agreement with the experimental observation as $\Delta E$ gradually decreases with $x$. The discrepancy can be due to consideration of only the total energy differences. Inclusion of various free energy contributions can make the deep energy minimum vanish, in agreement with the experiment. We see the same trend of gradual de-stabilisation of the martensitic phases with increasing concentration of the substituents for (Mn$_{2-x}$Co$_{x}$NiGa and Mn$_{2}$Ni(Ga$_{1-x}$Fe$_{x}$) systems. A comparison of the $\Delta E$ values for systems presented in Figs. \ref{tot-en}(a) and (c) indicate that Co-substituted systems have greater stability of the martensitic phases in comparison to Fe-substituted systems.

\begin{figure}[t]
\centerline{\hfill
\psfig{file=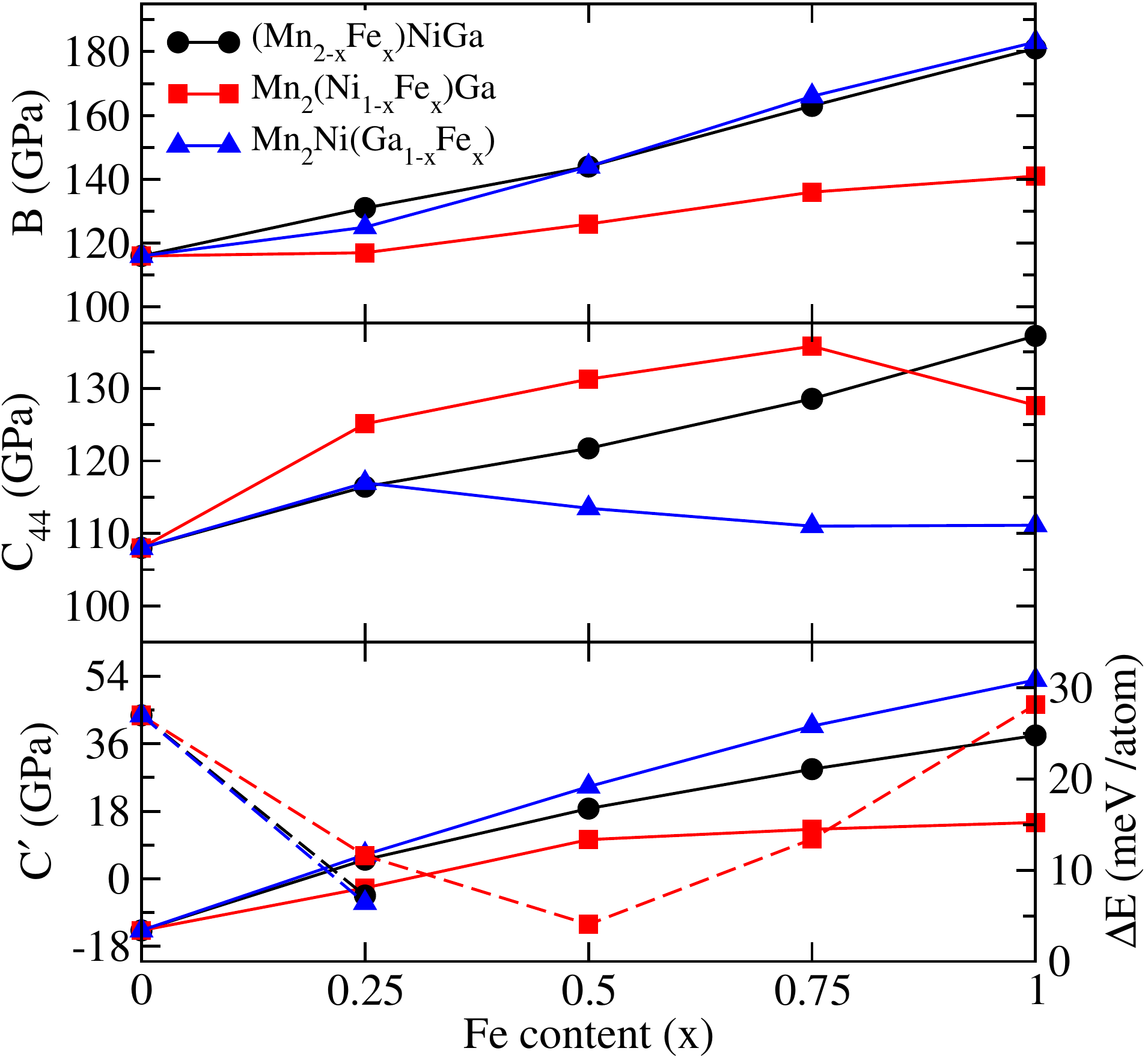,width=0.40\textwidth}\hfill}
\caption{The calculated bulk modulus(B), shear elastic constants
  C$_{44}$ and $C^{\prime}$ as a function of Fe content at different
  sites in Mn$_2$NiGa. The dashed lines represent variations of $\Delta{E}$ with $x$; $\Delta{E}$ is defined in Table~\ref{table1}.}
\label{fe-elas}
\end{figure}

In Fig. \ref{tot-en}(b) and (d), we compare the cases between Fe and Co substitutions, respectively, at the Ni sites. We find a gradual de-stabilisation of the martensitic phase with increasing $x$ for Mn$_{2}$(Ni$_{1-x}$Co$_{x}$)Ga system. A shallow minima at a (c/a) $\neq$ 1 for $x=0.5$ followed by absence of any minima for c/a $\neq$ 1 for higher $x$ indicate that martensitic transformation at reasonable temperatures might happen for upto $x=0.25$. This, once again, is slightly different from the experimental observations that no traces of martensitic transformation were obtained for $x>0.16$. \cite{MaPRB11} The case of Mn$_{2}$(Ni$_{1-x}$Fe$_{x}$)Ga is somewhat different from the rest. Here, we see gradual decrease in $\Delta E$ with $x$ indicating the gradual de-stabilisation of the martensitic phase upto $x=0.5$. However, as $x$ increases further, $\Delta E$ increases with the highest $\Delta E$ obtained for $x=1$, that is, for the compound Mn$_{2}$FeGa. An inspection of the relative volume changes in Table \ref{table1} shows that for this system, the relative changes in volume with $x$ are substantial and as large as $2.53 \%$ for Mn$_{2}$FeGa. Thus, although there is a cubic-to-tetragonal transformation, it is not volume conserving and thus is not martensitic as is the case for shape memory alloys. The physics of cubic-to-tetragonal phase transformation in Mn$_{2}$FeGa is very different as discussed in Ref. \onlinecite{Kundumn2fega17} and thus the trends observed in our calculations do not suggest any anomalous behaviour in this system.

\begin{figure}[t]
\centerline{\hfill
\psfig{file=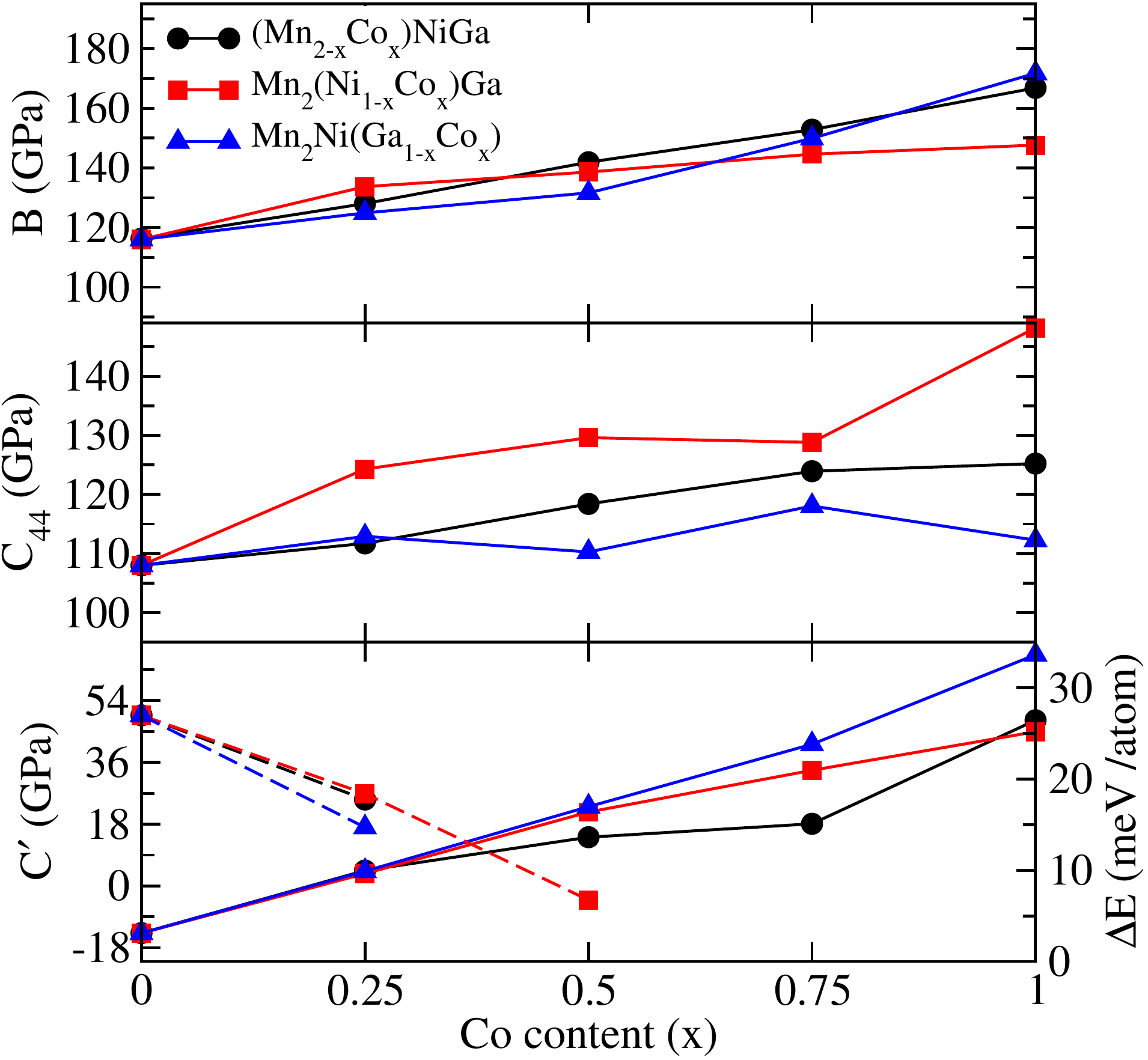,width=0.4\textwidth}\hfill}
\caption{The calculated bulk modulus(B), shear elastic constants
  C$_{44}$ and $C^{\prime}$ as a function of Co content at different
  sites in Mn$_2$NiGa. The dashed lines represent variations of $\Delta{E}$ with $x$; $\Delta{E}$ is defined in Table~\ref{table1}.}
\label{co-elas}
\end{figure}

The Fig. \ref{tot-en} and the tabulated (e/a) and $\Delta E$ values in Table \ref{table1} now suggest that the $\Delta E$ can be considered as a predictor for the qualitative variations in the T$_{m}$. However, in the systems under consideration, T$_{m}$ versus (e/a) correlation is absent except in Mn$_{2}$(Ni$_{1-x}$Co$_{x}$)Ga system, suggesting that (e/a) would not be a good predictor for T$_{m}$. This is consistent with the experimental results.\cite{LuoJAP10,MaPRB11} Summarising, we find a universal trend of de-stabilisation of the martensitic phase with increasing Fe or Co concentration, irrespective of the site at which they are substituted. The martensitic phases are stable mostly at low concentrations (upto $x=0.25$) of the substituent. In the next sub-sections we analyse the reasons behind this.

%========================================

\subsection{Elastic properties }

The composition dependent variations in the elastic constants of the high temperature austenite phase having cubic symmetry can often be a predictor of the martensitic transformation for the Ni-Mn-Ga system. \cite{ChunmeiPRB10,GhoshPBCM11} In Fe, Co and Cu substituted Ni$_{2}$MnGa, it has been established that the shear modulus $C^{\prime}$ can be a better predictor of composition dependent T$_{m}$ \cite{ChunmeiPRB11,ChakrabartiPRB13} than e/a since the martensitic transformations in these systems are related to the soft phonon modes which, in turn, are associated with the tetragonal shear elastic constant $C^{\prime}$ in the high temperature austenite phase. In Mn$_{2}$NiGa, the mechanism of martensitic phase transformation is quite similar to Ni$_{2}$MnGa.\cite{Kundumodulation17} So the calculation of $C^{\prime}$ along with other elastic constants would be useful to verify whether $C^{\prime}$ can be a good predictor of the composition dependence of martensitic transformation apart from the fundamental understanding of the composition dependence of the mechanical stability of these systems.

In Fig. \ref{fe-elas} and \ref{co-elas}, we present the results of calculated bulk modulus and shear modulii C$_{44}$ and $C^{\prime}$ for Fe and Co substituted Mn$_{2}$NiGa. For all  systems, the bulk modulus increases as the volume decreases which is consistent with the expected general trend. The elastic modulus C$_{44}$ is positive for all the alloys which satisfies one of the stability criteria for crystals with cubic symmetry. The value of $C^{\prime}$ increases with increasing of Fe or Co concentration which indicates that the system is increasingly insusceptible to a martensitic transformation. Negative or very low values of $C^{\prime}$ upto $25 \%$ of the substituent concentration for all systems are consistent with the obtained trends in the total energy minima with compositions(Fig. \ref{tot-en}). For Mn$_{2}$(Ni$_{1-x}$Fe$_{x}$)Ga system, we notice that the $C^{\prime}$ is almost constant beyond $x=0.25$ as opposed to the increasing trend (with $x$) for other systems. This is, once again, consistent with the fact that at all compositions in this particular system, the cubic phase transforms to a tetragonal phase with large energy cost, in particular for $x>0.5$, although such a phase transformation is not a volume conserving martensitic one.
The important outcome of the variations in the $C^{\prime}$ with compositions is that it can be considered as a better predictor of the composition dependence of T$_{m}$. In the previous sub-section, we have demonstrated that $\Delta E$ is a good predictor of the martensitic transformation. However, in order to fix a predictor, one need a physically measurable quantity. The justification of considering $C^{\prime}$ as the one can be understood by looking at the variations of $C^{\prime}$ with $\Delta E$ as shown in Fig. \ref{fe-elas} and \ref{co-elas}. The $C^{\prime}$ has a inverse relationship with $\Delta E$ which is according to our expectations; the former indicating increasing difficulty in de-stabilising the Hg$_{2}$CuTi structure against tetragonal shear while the later stands for the possibility of obtaining a martensitic transformation. Another significance of this result is that one can immediately find a similarity with regard to fixing  $C^{\prime}$ as the predictor of the composition dependence of T$_{m}$  in Fe, Cu and Co substituted Ni$_{2}$MnGa systems. To make sure that this is indeed true in cases of substitutions of all of these three atoms in Mn$_{2}$NiGa also, we have plotted the variations of $B, C_{44}$ and $C^{\prime}$ with concentration of the substituent for Cu substitution in various sites of Mn$_{2}$NiGa (Fig. 1, supplementary material) along with variations in $\Delta E$(values taken from Ref. \onlinecite{ChakrabartiPRB13}). We see the same inverse relationship between $C^{\prime}$ and $\Delta E$ in this system as well. Thus, we can conclude that $C^{\prime}$ can be considered as the predictor of the composition dependence of martensitic phase transformation in Ni-Mn-Ga alloy system, irrespective of substitution by another magnetic atom.

\begin{figure}[t]
\centerline{\hfill
\psfig{file=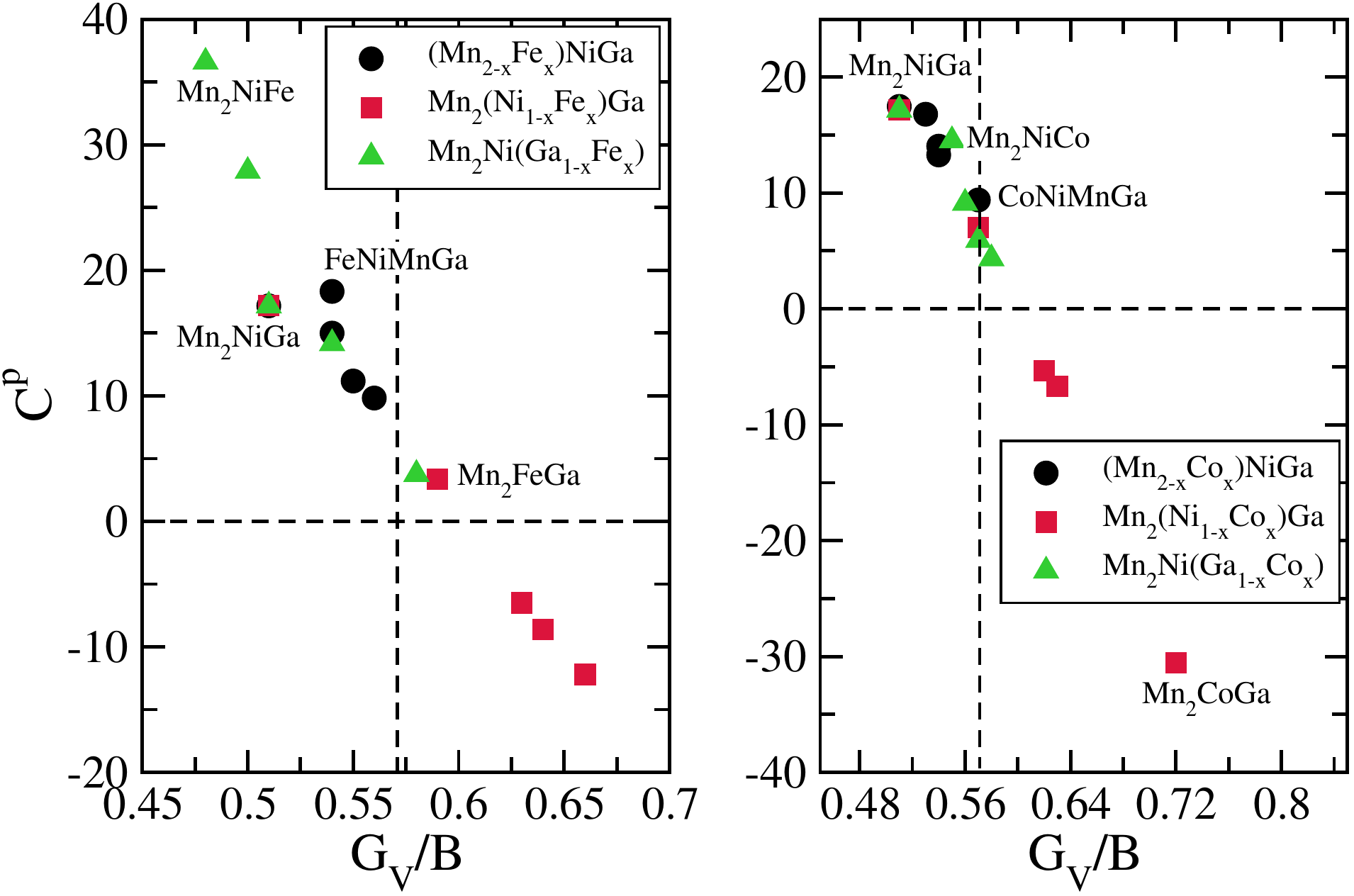,width=0.5\textwidth}\hfill}
\caption{Variations in the Cauchy pressure $C^{p}=(C_{12}-C_{44})$ with Pugh ratio $G_{v}/B$ are plotted for all the compounds. According to Ref. \onlinecite{NiuSR12}, $C^{p}>0, G_{v}/B<0.57$ indicates more ductility and more component of metallic bonding while $C^{p}<0, G_{v}/B>0.57$ indicates more brittleness and stronger covalent bonding.}
\label{cauchy-pugh}
\end{figure}

\begin{table*}[t]
\centering

\caption{\label{table2} Elastic constants of Co and Fe substituted Mn$_{2}$NiGa in their Hg$_{2}$CuTi cubic phases.}

\begin{tabular}{l@{\hspace{0.4cm}} c@{\hspace{0.4cm}} c@{\hspace{0.4cm}} c@{\hspace{0.4cm}} c@{\hspace{0.4cm}} c@{\hspace{0.4cm}} c@{\hspace{0.4cm}} c@{\hspace{0.4cm}} c@{\hspace{0.4cm}} }
\hline\hline
\vspace{-0.33 cm}
\\ System  & $B$ & $C^{\prime}$ &  $C_{44}$&  $C_{11}$ &  $C_{12}$  & $G_{v}$& $G_{v}/B$& $C^{p}=C_{12}-C_{44}$   \\
  & (GPa)&(GPa) & (GPa) &  (GPa) &  (GPa) & (GPa) &  & (GPa)   \\ \hline\hline

  Mn$_{2}$NiGa                     & 116 & -13.75 & 108 &  97.67 & 125.17 &  59.3 & 0.51 & 17.17 \\
  (Mn$_{0.75}$Fe$_{0.25}$)NiMnGa   & 131 & 5.1 & 116.43 & 137.8 & 127.6 &  71.9 &  0.55 &  11.17  \\
  (Mn$_{0.5}$Fe$_{0.25}$)NiMnGa    & 144 & 18.69 & 121.72 & 168.92& 131.54 &  80.51 &  0.56 &  9.82  \\
  (Mn$_{0.25}$Fe$_{0.75}$)NiMnGa   & 163 & 29.2 & 128.54 & 201.93& 143.53 &  88.8 &  0.54 &  14.99 \\
  FeNiMnGa                         & 181 & 38.18 & 137.23 & 231.91 & 155.55 &  97.61 & 0.54  &  18.32 \\
\hline
  Mn$_{2}$NiGa                        & 116 & -13.75 & 108 &  97.67 & 125.17 &  59.3 & 0.51 & 17.17    \\
  Mn$_{2}$(Ni$_{0.75}$Fe$_{0.25}$)Ga   & 117 & -2.41 & 125.1 & 113.79 & 118.61 & 74.1  & 0.63  & -6.49   \\
  Mn$_{2}$(Ni$_{0.5}$Fe$_{0.5}$)Ga     & 126 & 10.42   & 131.25 & 139.89  & 119.05 & 82.92  & 0.66  & -12.2   \\
  Mn$_{2}$(Ni$_{0.25}$Fe$_{0.75}$)Ga   & 136 & 13.2 & 135.8 & 153.6 & 127.2 & 86.76  & 0.64  &  -8.6  \\
  Mn$_{2}$FeGa                         & 141 & 15  & 127.64 & 161 & 131 & 82.58  & 0.59 &  3.36  \\
\hline
 Mn$_{2}$NiGa                          & 116 & -13.75 & 108 &  97.67 & 125.17 &  59.3 & 0.51 & 17.17  \\
  Mn$_{2}$Ni(Ga$_{0.75}$Fe$_{0.25}$)   & 125 & 6.44 & 116.96 & 133.59 & 120.71 & 72.75  &  0.58 & 3.75 \\
  Mn$_{2}$Ni(Ga$_{0.5}$Fe$_{0.5}$)     & 144 & 24.56  & 113.46 & 176.75& 127.63  &  77.9 &  0.54 & 14.17 \\
  Mn$_{2}$Ni(Ga$_{0.25}$Fe$_{0.75}$)   & 166 & 40.66 & 111 & 220.21 & 138.89  & 82.86 & 0.5  & 27.89 \\
  Mn$_{2}$NiFe                         & 183 & 52.94     & 111.12 & 253.59 & 147.71  & 87.85  &  0.48 &  36.59\\
\hline
   Mn$_{2}$NiGa                          & 116 & -13.75 & 108 &  97.67 & 125.17 &  59.3 & 0.51 & 17.17 \\
  (Mn$_{0.75}$Co$_{0.25}$)NiMnGa       & 128 & 4.5 & 111.71 & 133.99 & 125  &  68.82 &  0.54 & 13.29 \\
  (Mn$_{0.5}$Co$_{0.25}$)NiMnGa        & 142 & 14.21     & 118.39 & 160.84 & 132.43&  76.72 & 0.54   & 14.04 \\
  (Mn$_{0.25}$Co$_{0.75}$)NiMnGa       & 153 & 18.11 & 123.93 & 176.94 & 140.73  & 81.6 & 0.53  & 16.8 \\
  CoNiMnGa                             & 167 & 48.27  & 125.22 & 231.17 & 134.62 &  94.44 & 0.57 & 9.4 \\
\hline
  Mn$_{2}$NiGa                         & 116 & -13.75 & 108 &  97.67 & 125.17 &  59.3 & 0.51 & 17.17  \\
  Mn$_{2}$(Ni$_{0.75}$Co$_{0.25}$)Ga   & 133.7 & 3.61 & 124.28 & 138.51 & 131.29  & 76.01  &  0.57 & 7.01 \\
  Mn$_{2}$(Ni$_{0.5}$Co$_{0.5}$)Ga     & 138.6 & 21.51  & 129.61 & 167.28 & 124.26  & 86.37  &  0.62 & -5.35 \\
  Mn$_{2}$(Ni$_{0.25}$Co$_{0.75}$)Ga   & 144.6 & 33.7 & 128.81 & 189.54 & 122.13  & 90.77  & 0.63  &  -6.38\\
  Mn$_{2}$CoGa                         & 147.6 & 44.93    & 148.18 & 207.51 & 117.65 &  106.88 & 0.72  & -30.53 \\
\hline
  Mn$_{2}$NiGa                         & 116 & -13.75 & 108 &  97.67 & 125.17 &  59.3 & 0.51 & 17.17  \\
  Mn$_{2}$Ni(Ga$_{0.75}$Co$_{0.25}$)   & 124.9 & 4.3 & 112.9 & 130.63 & 122.03  & 69.46  &  0.56 & 9.13 \\
  Mn$_{2}$Ni(Ga$_{0.5}$Co$_{0.5}$)     & 131.6 & 23.09  & 110.28 & 162.38& 116.21 & 75.4  &  0.57 & 5.93 \\
  Mn$_{2}$Ni(Ga$_{0.25}$Co$_{0.75}$)   & 149.9 & 41.29 & 118.04 & 204.95 & 122.37  & 87.34  &0.58   & 4.33 \\
  Mn$_{2}$NiCo                         & 171.7 & 67.36  & 112.26 & 261.52 & 126.79   &  94.3 &  0.55 & 14.53 \\
              
\hline\hline
\end{tabular}
%}
\end{table*}

One of the reasons for substituting another magnetic atom into Mn$_{2}$NiGa is to improve it's mechanical properties, such as ductility. A good measure of whether the system is more ductile or more brittle is to look at it's Pugh ratio \cite{PughPM54} given as $G_{v}/B$, $G_{v}$ the isotropic shear modulus under Voigt formalism \cite{VoigtAP89} related to the resistance of the material to plastic deformation and $B$ the Bulk modulus. A $G_{v}/B$ value of 0.57 is considered critical to decide on the brittleness of the compound. Compounds having the Pugh ratio greater than 0.57 are considered more brittle. On the other hand, Cauchy pressure, defined as $C^{p}=(C_{12}-C_{44})$ provides an insight to the nature of bonding in a  material with cubic symmetry. \cite{PettiforMST92} According to this, a positive value of Cauchy pressure indicates presence of more metallic bonding in the system while a negative value implies stronger component of covalent bonding. Very recently, Niu \textit{et al.} has shown that the Pugh ratio and Cauchy pressure are well correlated with their ductile-to-brittle transition, matching with the metallic-to-covalent bonding transformation for a number of cubic crystals.\cite{NiuSR12} In Fig. \ref{cauchy-pugh}, we plot the Pugh ratio versus Cauchy pressure for all the compounds studied here. For better understanding of the trends for each type of substitution, the values of the elastic modulii are given in Table \ref{table2}. The results as displayed in Fig. \ref{cauchy-pugh} imply that a correlation between $C^{p}$ and $(G_{v}/B)$ as suggested in Ref. \onlinecite{NiuSR12} and is seen in case of a group of Co$_{2}$ and Ni$_{2}$-based Heusler alloys. \cite{RoyPRB16} However, the absolute numbers in Table \ref{table2} suggest that for a given group of systems, that is, systems obtained by gradual substitution of a particular atom at a particular site of Mn$_{2}$NiGa, this correlation does not necessarily hold. For example, one can see a linear variation between $C^{p}$ and $(G_{v}/B)$ in (Mn$_{2-x}$Fe$_{x}$)NiGa for $x \leq 0.5$ after which the $C_{p}$ increases with almost no variation in $(G_{v}/B)$. Overall, the results imply that the systems are very close to ductile-brittle critical limit and that the bonding has more metallic component, the two notable exceptions are when Co or Fe is substituted at the Ni site. As the Ni content in the systems decrease, we observe the tendency of the system to be more brittle and the bonding having more covalent component. Mn$_{2}$CoGa has the highest $(G_{v}/B)$ ratio of 0.72 which is close to Si and Ge \cite{NiuSR12} along with a large negative $C^{p}$.  The effect is not as dramatic in case of Mn$_{2}$(Ni$_{1-x}$Fe$_{x}$)Ga although a high value of $(G_{v}/B)$, 0.66, along with a $C^{p}$ value -12.2 are obtained for $x=0.5$. Beyond $x=0.5$, the system tends to be more metallic and ductile although the numbers for $x=0.75$ and $1$ are closer to the transition lines. Thus, the bonding features in cases of substitutions at Ni site must be different from the rest.

\subsection{Electronic structure}
\begin{figure}[t]
\centerline{\hfill
\psfig{file=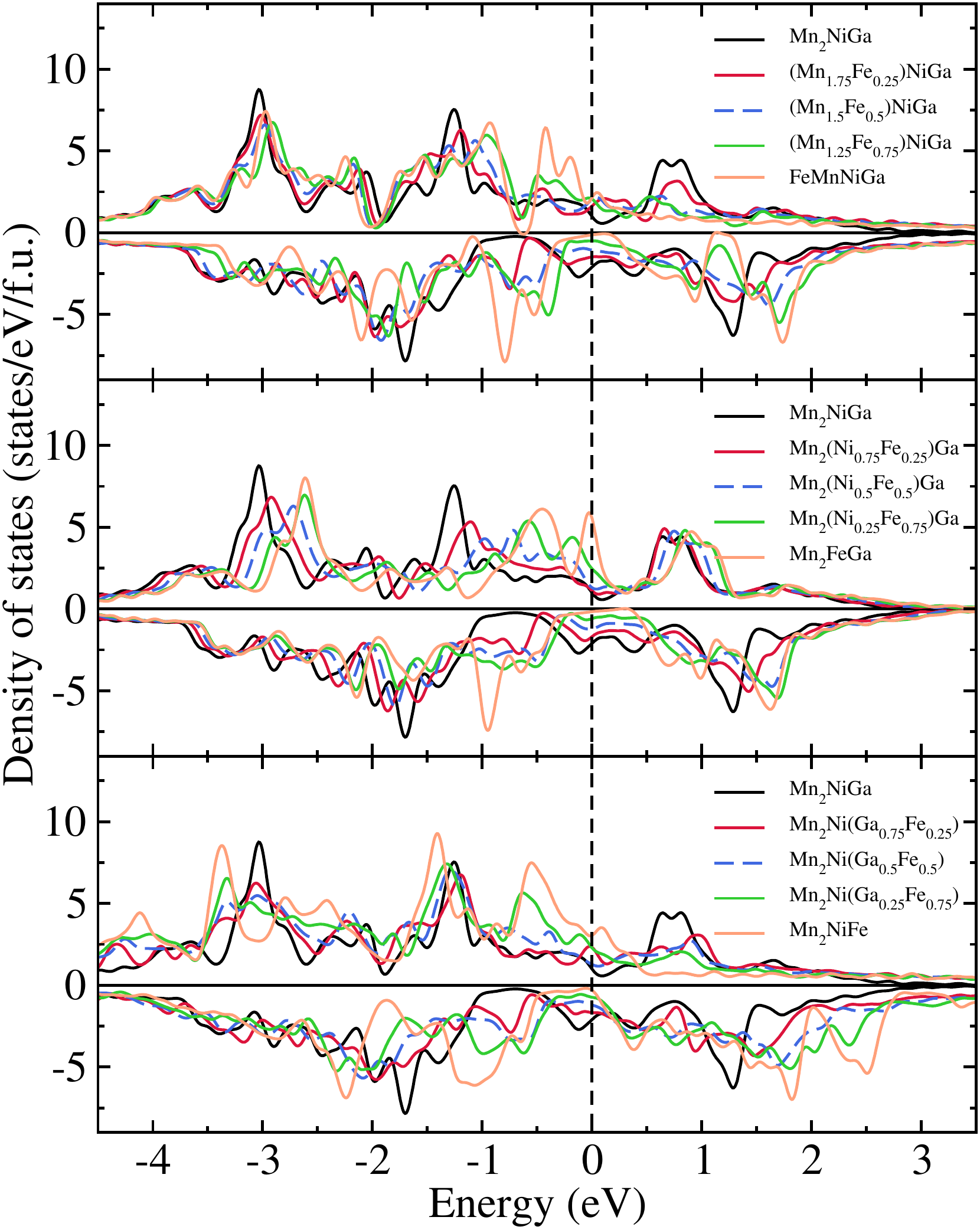,width=0.4\textwidth}\hfill}
\caption{Total density of states for Fe substituted at different sites in
  Mn$_2$NiGa. The zero of the energy is set at Fermi energy(E$_{F}$).}
\label{fe-dos}
\end{figure}

\begin{figure}[t]
\centerline{\hfill
\psfig{file=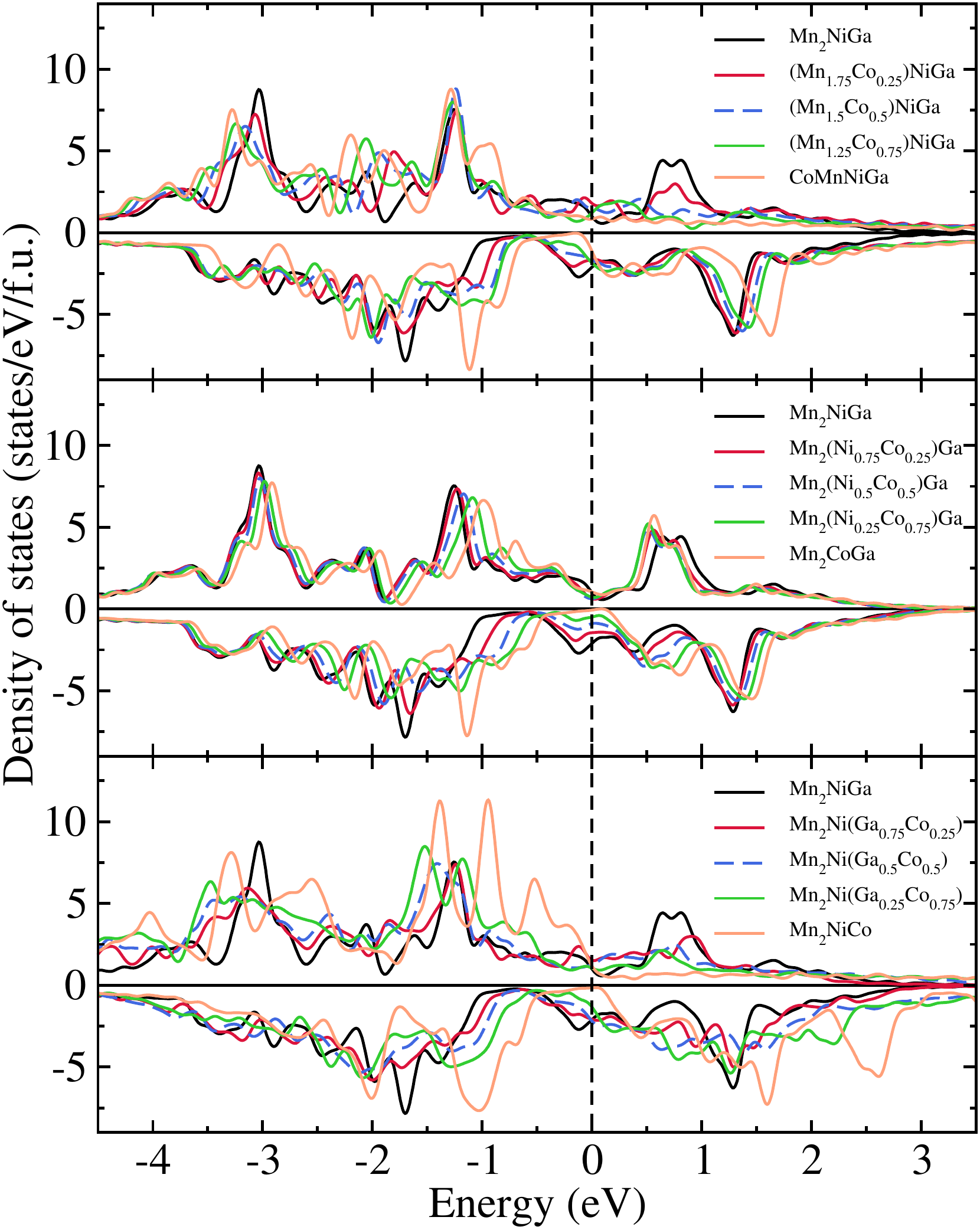,width=0.4\textwidth}\hfill}
\caption{Total density of states for Co substituted at different sites in
  Mn$_2$NiGa. The zero of the energy is set at Fermi energy(E$_{F}$).}
\label{co-dos}
\end{figure}

The results in the previous two sub-sections demonstrate that the martensitic phase gets gradually de-stabilised as the concentration of the substituent in any one of the sites of Mn$_{2}$NiGa increases. For most of the systems investigated, a concentration (of the substituent) beyond $25 \%$ leads to a complete stabilisation of the cubic austenite phase. At the same time, one sees that except substitutions at the Ni site, the bonding remains dominantly metallic in nature. In this sub-section we try to understand the reasons behind these trends from the composition dependent variations of the electronic structures in the substituted Mn$_{2}$NiGa systems. To do this, we have plotted the total densities of states of compounds where Fe and Co are substituted in various sites. The results for Fe substitution at different sites are shown in Fig. \ref{fe-dos} and those for Co-substitution at different sites are shown in Fig. \ref{co-dos}.

In the Hg$_{2}$CuTi phase of pristine Mn$_{2}$NiGa, there exists a pseudogap in the minority spin band at about 1 eV below the Fermi level. This pseudogap is formed mainly due to the hybridisations between the $3d$ states of MnI and Ni which occupy symmetric positions in the Hg$_{2}$CuTi lattice and the $4p$ states of Ga. A peak around 0.1 eV below Fermi level, originating from the hybridisations of the same orbitals results in the Jahn-Teller instability in the system \cite{BarmanPRB08,PaulJAP11,PaulJAP14,Kundumodulation17} and drives the system towards the martensitic transformations. The stabilities of the martensitic phases as obtained from the total energy calculations, the variations in the elastic modulii and proposed changes in the strengths of bonding upon substitutions can now be interpreted from the minority spin band densities of states. The elastic constant $C^{\prime}$ is directly connected to the Jahn-Teller distortion since it is the elastic modulus of tetragonal deformation. Thus, a stronger covalent bond or weaker Jahn-Teller distortion should result in a harder $C^{\prime}$.

Figs. \ref{fe-dos} and \ref{co-dos} clearly show that substitutions weaken the Jahn-Teller distortion in Mn$_{2}$NiGa as the peak around -0.1 eV gradually gets smeared out, thus stabilising the Hg$_{2}$CuTi structure. This explains the gradual de-stabilisation of the martensitic phases with concentrations of the substituents as seen in Figs. \ref{tot-en}. Irrespective of the site at which the substitution is done, it is either MnI or Ni content that decreases at the 4a and 4b sites weakening the hybridisations between their $3d$ and Ga $4p$ states. The substituents, either Co or Fe cannot restore the hybridisation as their $3d$ states in the minority bands lie much deeper (Figs. 2-3 in supplementary information). Such weakening of the Jahn-Teller distortion upon substitutions is the reason behind hardening of $C^{\prime}$ with increasing substituent content in Mn$_{2}$NiGa.

As seen from Figs. \ref{fe-dos} and \ref{co-dos} as well as Figs. 2 and 3 of supplementary material, for the Mn$_{2-x}$X$_{x}$NiGa (X=Fe, Co) systems, the pseudogap at about -1 eV in the minority bands  gradually become narrower and shallower for X=Fe. This happens primarily due to the position of the Fe-$d$ states which are right in the gap. The weakening of the Jahn-Teller distortion gradually pushes the MnI and Ni states towards lower energies, making them hybridise with Fe states within 0.3 eV to 0.75 eV below Fermi level. Thus, initially for $x \leq 0.5$, the covalent bond strength in the system increases and is reflected in the changes in $C^{p}$ (Table \ref{table2}). Beyond $x=0.5$, the MnI content reduces weakening the covalent bonding slightly($C^{p}$ values increase again along with a decrease in the $(G_{v}/B$)).  In case of X=Co, the situation is slightly different. The Co states lie deeper than Fe states, but the hybridisations with Ni and MnI states which are pushed in to the pseudo-gap due to weakening of the Jahn-Teller effect hybridise with the Co states in a similar way as Fe. The only noticeable difference in the two cases is that while the Fe states are more delocalised towards higher energies, the Co states are localised strengthening the covalent components in the bonding. In both cases the pseudo-gap moves towards higher energies but in case of the Fe substitution it morphs into a real gap cutting through the Fermi level. A look at the majority spin densities of states can also help in understanding the evolution of the bonding strengths. For the Fe-substituted systems, more states pile up near the Fermi level when the $x>0.5$ due to increased hybridisation of MnII, Fe and Ga states contributing to more metallicity in the bonds. In case of Co-substitution at the MnI site, there is very little changes in the majority band densities of states and thus the evolution in the bonding nature is to be understood from the features in the minority band densities of states.

In case of Mn$_{2}$Ni(Ga$_{1-x}$Fe$_{x}$), the minority band densities of states are dominated by Fe states  weakly hybridising with other atoms. The Fe states are more delocalised except for $x=0.25$. However, MnI and Ni hybridise strongly for $x=0.25$ with the strength gradually diminishing. The majority band densities of states have noticeable changes near the Fermi level with increasing $x$ and beyond $x=0.25$; the major highlight being a peak and larger densities of states at Fermi level owing to hybridisations of Fe and MnIII states. This explains the more covalent nature of the bonds for $x=0.25$ and more metallic for higher $x$. In case of Mn$_{2}$Ni(Ga$_{1-x}$Co$_{x}$), there are little changes in the majority spin densities of states upto $x=0.75$. In the minority band, Co has greater contributions but it hybridises weakly with other atoms while the stronger hybridisations are between Ni and MnI in the intermediate concentrations. This means that the covalent bonding gets strengthened supporting the results of Table \ref{table2}. The absence of MnI atoms at $x=1$ along with sharp contributions from MnIII near the Fermi level in the majority spin band, weakens the covalent bond strengths and makes the system more metallic.

From Table \ref{table2}, it appears that the substitutions at Ni sites makes the covalent bonds stronger and it prevails for Co-substitution, in particular. For Co-substituted systems, we find that that there is very little changes in the overall features of the majority band densities of states as the Co-content increases. The major changes occur in the minority spin band where Co, MnI and Ni hybridise. As the Ni content decreases, the hybridisations between Co and MnI (in the range of -1.25 eV to -0.5 eV) gets stronger making the character of the bonds more covalent. In case of Fe-substitution the stronger covalent component in their bonds can be explained in a similar way upto $x=0.5$. However, at $x=0.75$, a peak appears close to the Fermi level due to Fe-MnI hybridisation, which transforms to a large peak exactly at the Fermi level for $x=1$, thus increasing the metallic contributions to the bonds. This picture is consistent with the results from Table \ref{table2}.

\subsection{Total and atomic magnetic moments}

\begin{figure}[t]
\centerline{\hfill
\psfig{file=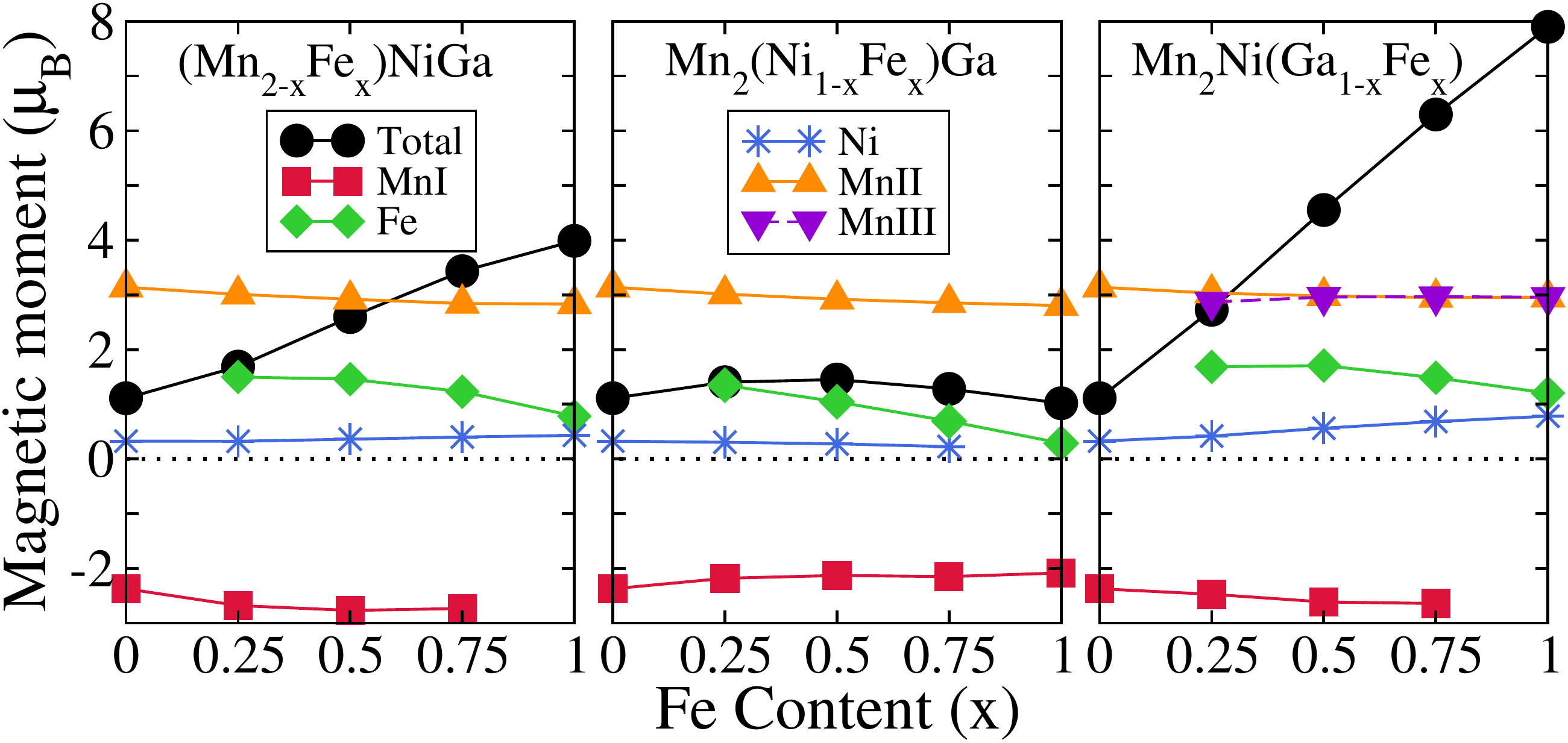,width=0.45\textwidth}\hfill}
\caption{The calculated total ($\mu_{\rm B}/f.u.$) and atomic magnetic moment as a function
  of Fe content for Fe-substituted Mn$_2$NiGa.}
\label{fe-moment}
\end{figure}

\begin{figure}[t]
\centerline{\hfill
\psfig{file=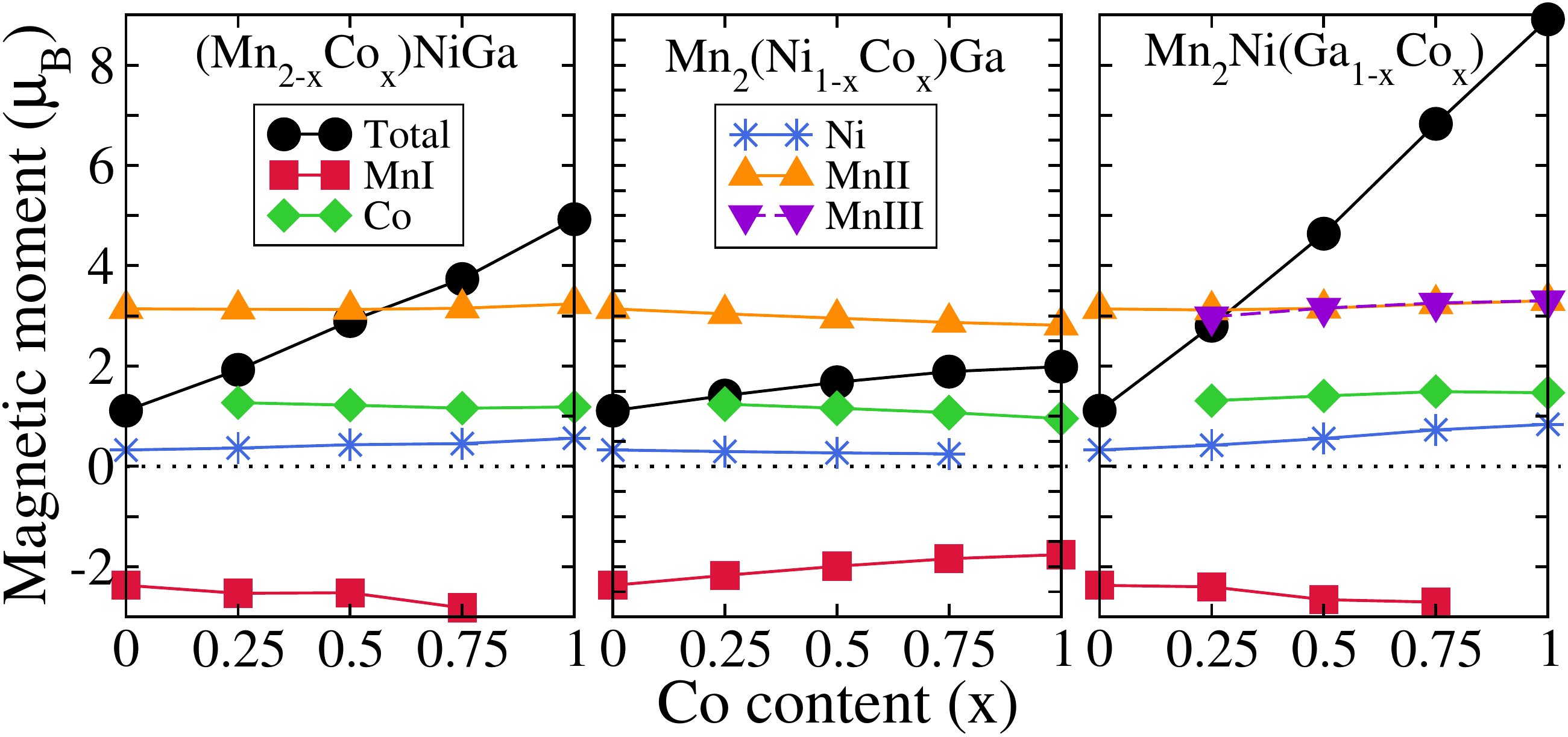,width=0.45\textwidth}\hfill}
\caption{The calculated total ($\mu_{\rm B}/f.u.$) and atomic magnetic moment as a function
  of Co content for Co-substituted Mn$_2$NiGa.}
\label{co-moment}
\end{figure}

In Table \ref{table1} and Figs. \ref{fe-moment} and \ref{co-moment} we present our results on total and atomic magnetic moments to understand the effects of site-substitution in Mn$_{2}$NiGa in the Hg$_{2}$CuTi phase. Magnetisation measurements on (Mn$_{2-x}$Fe$_{x}$)NiGa, \cite{LuoJAP10} Mn$_{2}$(Ni$_{1-x}$Co$_{x}$)Ga and Mn$_{2}$Ni(Ga$_{1-x}$Co$_{x}$) \cite{MaPRB11} systems showed an increase in magnetisation upon Fe and Co substitutions. From our results, we summarise the observations on the total magnetic moments: (i) In almost all cases the total moment increases with the concentration of the substituent, (ii) the rise in the total moment is fastest for Mn$_{2}$Ni(Ga$_{1-x}$X$_{x}$), and is slowest for Mn$_{2}$(Ni$_{1-x}$X$_{x}$)Ga systems, (iii) the variation of total moment with $x$ is non-linear for Mn$_{2}$(Ni$_{1-x}$Fe$_{x}$)Ga. It increases for $x \leq 0.5$ and then decreases for higher values of $x$. However, this is consistent with the substantial changes in the electronic structures discussed in the previous sub-section, (iv) all the compounds formed by complete substitution of one of the atoms in the parent compound ($x=1$) have near integer moments with the highest being $\sim 9 \mu_{B}$ for Mn$_{2}$NiCo. 

The calculated results agree very well with the ones available in the literature, obtained either in the magnetic measurements \cite{MaPRB11,LuoJAP10} or from the first-principles calculations. \cite{ChakrabartiPRB13,WollmannPRB14,AlijaniPRB11,BarmanEPL07} For (Mn$_{2-x}$Fe$_{x}$)NiGa ($x=0.25,0.5$) compounds, the calculated results of magnetic moments are consistent with the experiment.\cite{LuoJAP10} The noticeable disagreements are in cases of Mn$_{2}$Ni(Ga$_{1-x}$Co$_{x}$)($x=0.25,0.5$)\cite{MaPRB11} compounds. This could be because of presence of anti-site disorder in the experimental sample or due to the differences between actual composition and the one reported in the experiment. \cite{MaPRB11} Both the effects can affect the magnetic moment as it is found to be very sensitive to the sub-lattice composition in Ni-Mn-Ga systems. \cite{ChunmeiPRB11}

The changes in the compositions at various sites affect the atomic moments which in turn affect the variations in the total magnetic moments with the concentration of the substituents. From Figs. \ref{fe-moment} and \ref{co-moment}, we find the following trends: (i) like Mn$_{2}$NiGa, MnI and MnII atoms couple anti-parallely in all compounds where MnI is present (ii) the moment of MnII atoms undergoes little variation with the composition across all compounds; the general trend being a small decrease with the concentration of the substituent, (iii) the moment associated with the Ni atoms increase with concentration of the substituent except when it is substituted by Fe or Co; the faster increase is in cases of substitutions at Ga sites, (iv) the MnI magnetic moments increase slightly in magnitude making the antiferromagnetic component in the system stronger as the concentration of the substituent increases for all systems except when substitutions are done at Ni sites, where it's magnitude decreases signifying increase in the ferromagnetic component in the system; the change is more rapid in case of Co substitution at Ni site, (v) for the Ga-substituted systems, MnIII, the Mn atoms at the Ga sites have large ferromagnetic moments, almost equal in magnitude of MnII moments; as the concentration of the substituent increases the concentration of MnI(MnIII) decreases(increases) weakening the antiferromagnetic component in favour of a strong ferromagnetic one. This explains the rapid increase of the total moment for these systems, (vi) the Co moment is larger than Ni moment and undergoes little change across compositions and the site of substitution while the Fe moment has a general trend of decreasing with increasing Fe concentration. A rapid  quenching of Fe moment is observed in case of Mn$_{2}$FeGa, the reason of which has been discussed earlier. \cite{Kundumn2fega17} The slow increase in the total moment for Ni-substituted systems, thus, can be understood in terms of the losses of MnII and MnI moments concurrently but in opposite directions, leaving the changes to be governed solely by Ni and the substituents moments, both of which vary slowly across compositions. The linear increase in the total moment of Mn-substituted system, on the other hand, can be attributed to the gradual loss of contributions of MnI due to it's decreasing concentration which boosts the ferromagnetic component in the system with concentration of the substituent. The explanations on the nature of such variations in the atomic moments require analysis of the electronic structures of each constituent. This has been done in the supplementary material. 
%\clearpage
\subsection{Magnetic Exchange interactions and Curie temperature}

\begin{figure}[t]
\centerline{\hfill
\psfig{file=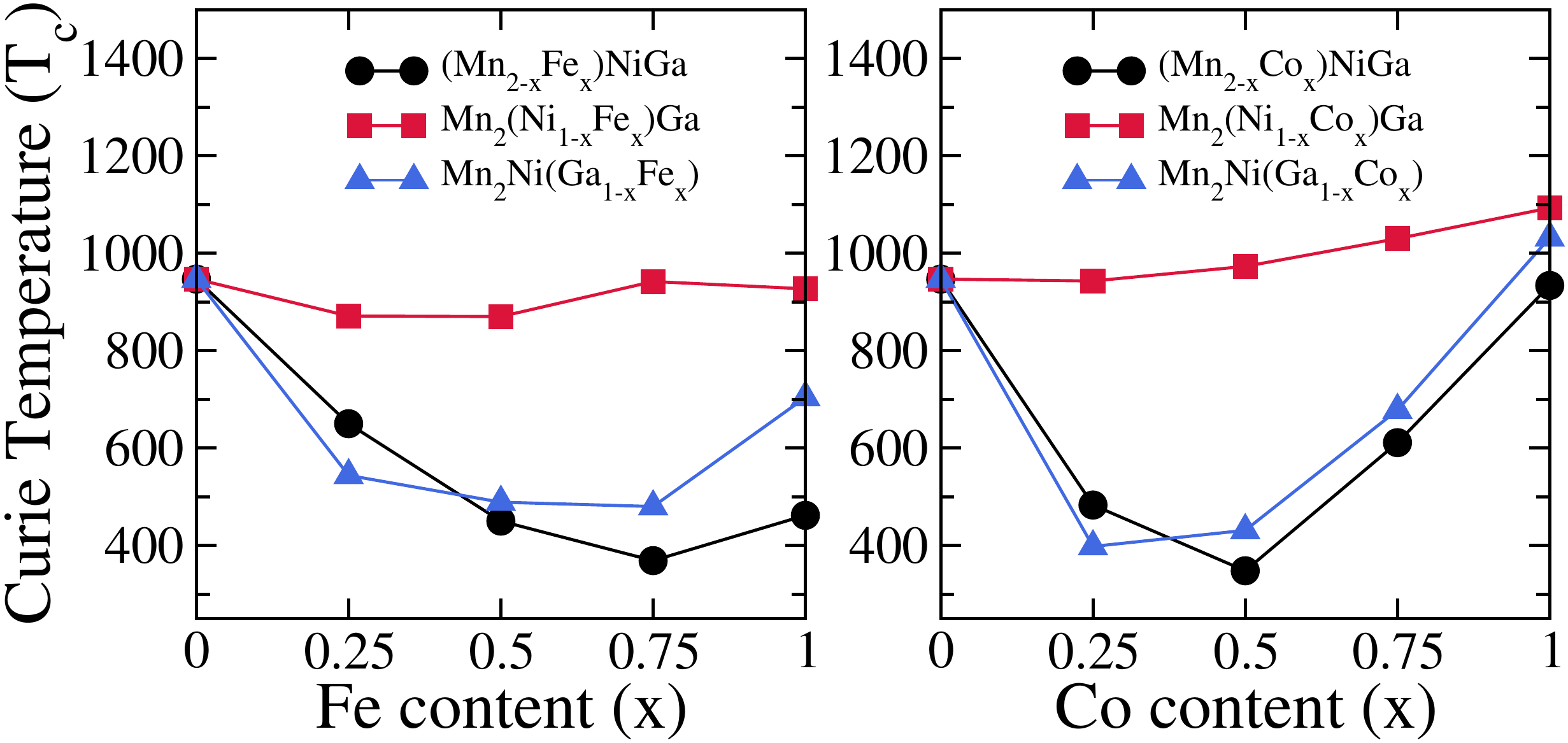,width=0.45\textwidth}\hfill}
\caption{Calculated Curie temperatures as a function of substituent (a) Fe concentration and (b) Co concentration
for substitutions at different sites in Mn$_2$NiGa. Calculations are done by MFA method.}
\label{mf-tc}
\end{figure}

\begin{figure}[t]
\centerline{\hfill
\psfig{file=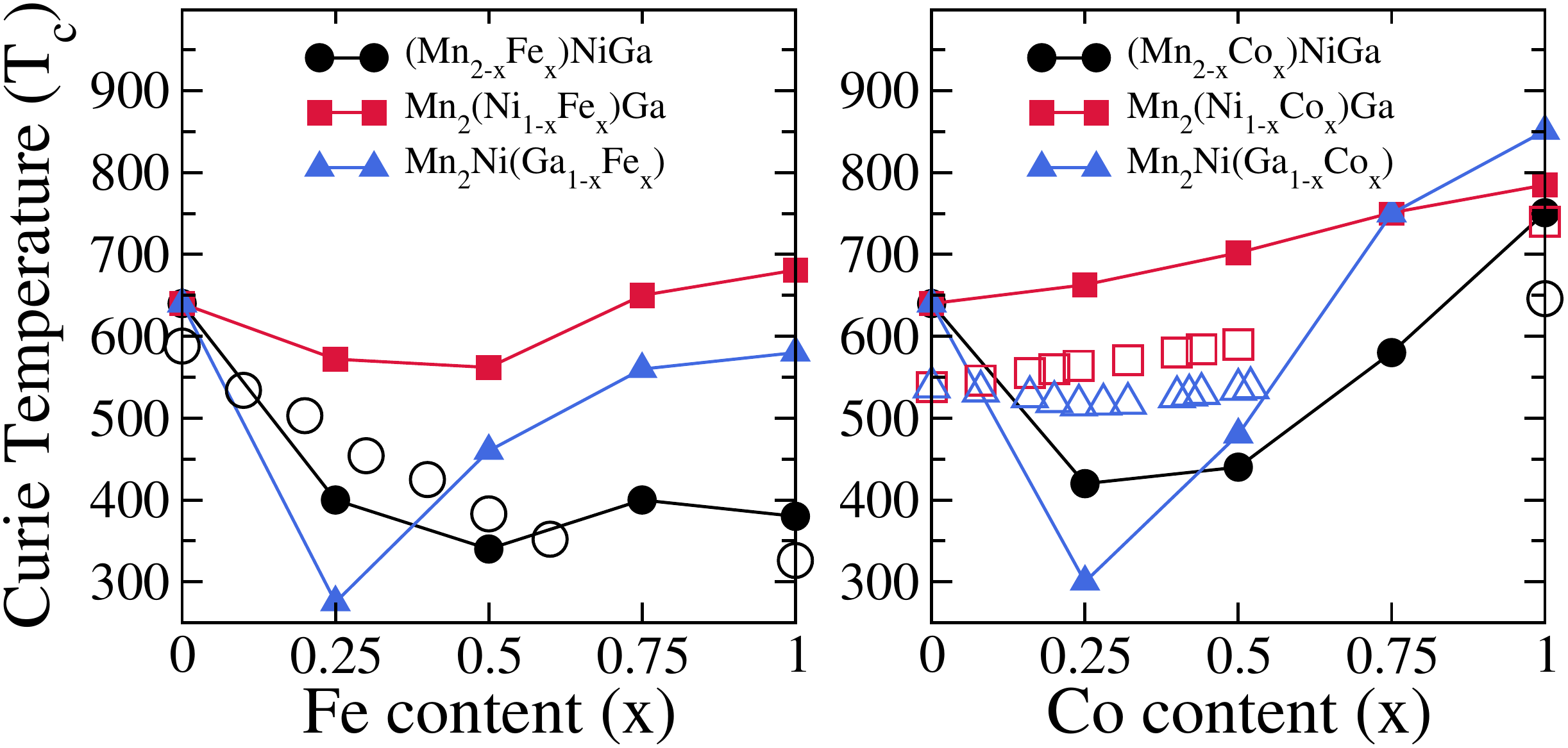,width=0.45\textwidth}\hfill}
\caption{Calculated Curie temperatures as a function of substituent (a) Fe concentration and (b) Co concentration
for substitutions at different sites in Mn$_2$NiGa. Calculations are done by MCS method. Open symbols represent the experimental results which are adopted from Ref. \onlinecite{MaPRB11,LuoJAP10,AlijaniPRB11,MinakuchiJAC15}.}
\label{mc-tc}
\end{figure}

In Fig. \ref{mf-tc} and \ref{mc-tc}, we show the  composition dependences of Curie temperatures (T$_{c}$) calculated using Mean field approximation (MFA) and Monte Carlo simulation (MCS), respectively, for Fe and Co substituted Mn$_{2}$NiGa. Since the MFA results typically overestimate the T$_{c}$ and MCS results are found to be closer to experimental results for variety of systems, the available experimental results in these compounds are included in Fig. \ref{mc-tc}. The trends in variations of T$_{c}$ calculated using MFA and MCS are by and large similar. As expected, the MFA results are over-estimated in comparison with the MCS results and the experimental values. From the MCS results, we find that all the end compounds obtained by complete substitution of Fe or Co in any of the sites have very high T$_{c}$, the largest being close to 900K for Mn$_{2}$NiCo. The qualitative variation of T$_{c}$ for Mn$_{2}$(Ni$_{1-x}$Co$_{x}$)Ga agrees well with the experiment \cite{MaPRB11,MinakuchiJAC15} although the MCS results are little overestimated. The quantitative agreement between the MCS results and the experimental results \cite{LuoJAP10,AlijaniPRB11} for (Mn$_{2-x}$Fe$_{x}$)NiGa are better. There is significant disagreement, both qualitatively and quantitatively for Mn$_{2}$Ni(Ga$_{1-x}$Co$_{x}$) systems. The MCS calculations show a sharp decrease in T$_{c}$ for $25\%$ of Co substitution after which the T$_{c}$ rises sharply with increase in $x$. The experiment, \cite{MaPRB11} on the other hand, although obtained a decrease in T$_{c}$ upto $x=0.24$ and a rise after that, the changes were not this substantial. The experimentally obtained T$_{c}$ decreased from 538K ($x=0$) to 517K for $x=0.24$ and rose to only 537K for $x=0.52$. It may be noted that their T$_{c}$ for Mn$_{2}$NiGa is lower by 50K in comparison to T$_{c}$ obtained from other experiments. \cite{LiuAPL05,LuoJAP10} Such discrepancies could be due to anti-site disorder or off-stoichiometric compositions present in the samples used by the authors of Ref. \onlinecite{MaPRB11}. In fact, we have reported discrepancies in the magnetic moments calculated by us and obtained from their magnetic measurements in the previous sub-section. Thus the origin of these discrepancies could be the same. Nevertheless, the agreement between the MCS results and the experiments for the end compounds, wherever available, is remarkable.

\begin{figure}[t]
\centerline{\hfill
\psfig{file=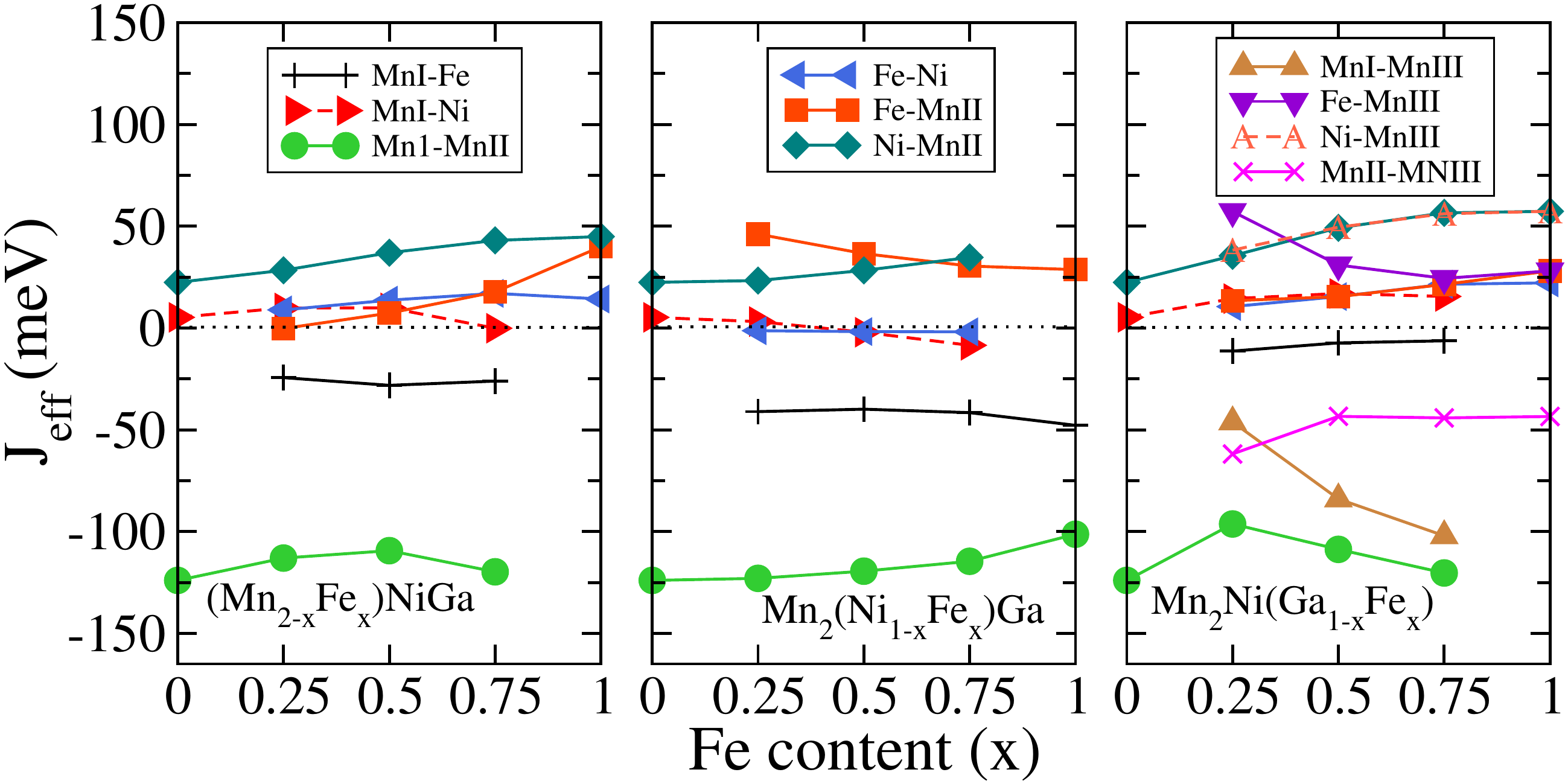,width=0.48\textwidth}\hfill}
\caption{Effective exchange coupling constant (J$_{eff}$) as a function of
  Fe content at different sites in Mn$_2$NiGa.}
\label{fe-exc}
\end{figure}

\begin{figure}[t]
\centerline{\hfill
\psfig{file=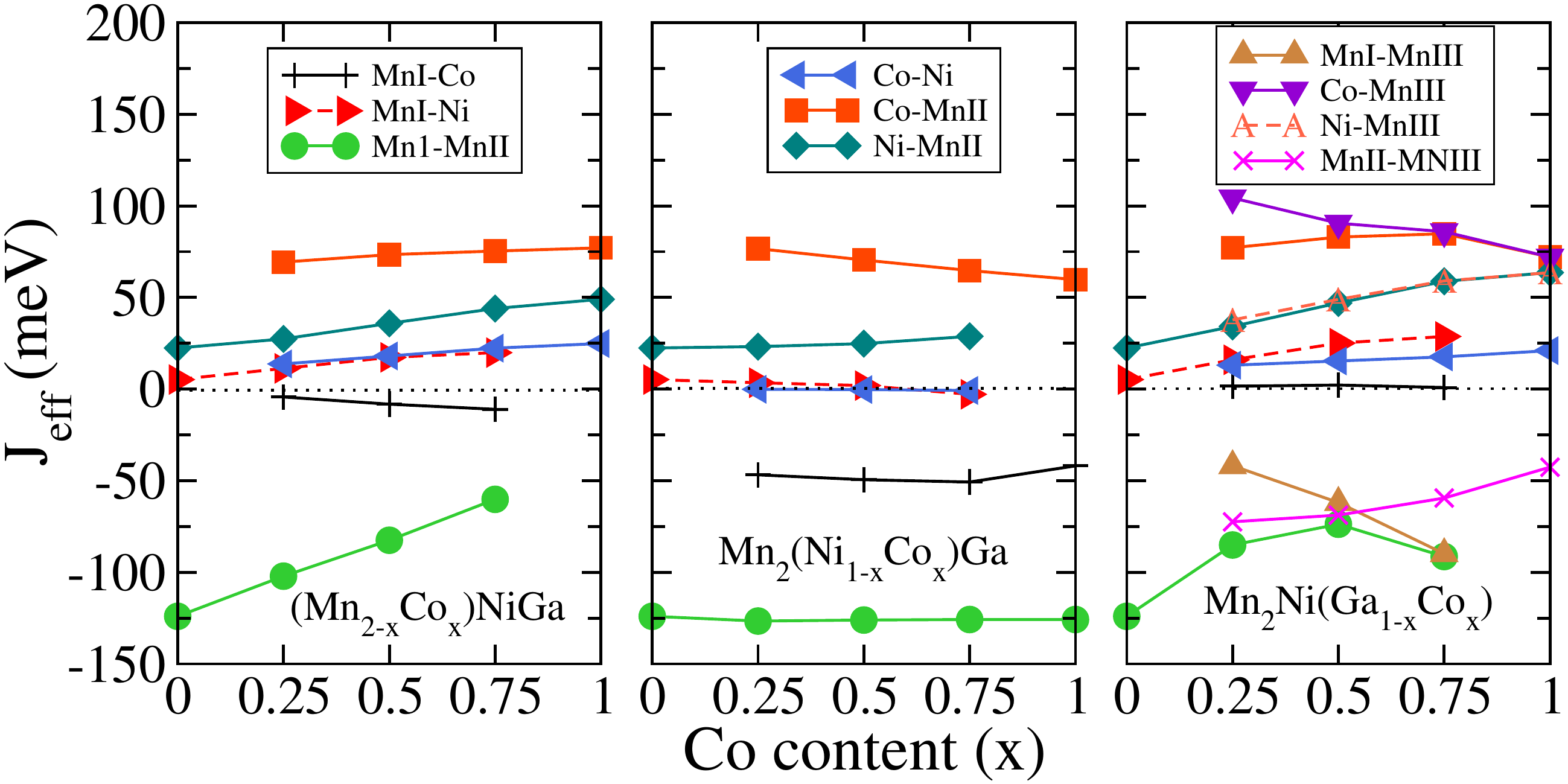,width=0.48\textwidth}\hfill}
\caption{Effective exchange coupling constant (J$_{eff}$) as a function of
  Co content at different sites in Mn$_2$NiGa.}
\label{co-exc}
\end{figure}

In order to understand the trends in the Curie temperature, we have calculated the inter-atomic magnetic effective exchange coupling(J$_{eff}$), presented in Fig. \ref{fe-exc} and \ref{co-exc} for Fe and Co substituted Mn$_{2}$NiGa, respectively. The J$_{eff}$ are calculated as $J_{eff}^{\mu\nu}=\sum_{j}J_{0j}^{\mu\nu}$; $0$ fixed to sub-lattice $\mu$ and the sites $j$ belong to sub-lattice $\nu$. We find that the dominant J$_{eff}$s remain either antiferromagnetic or ferromagnetic across compositions. The dominant antiferromagnetic J$_{eff}$ is due to the MnI-MnII pairs while the other dominant inter-sublattice J$_{eff}$ s are ferromagnetic. For (Mn$_{2-x}$Co$_{x}$)NiGa system, the strength of antiferromagnetic J$_{eff}^{MnI-MnII}$ decreases with $x$ while the strength of the largest ferromagnetic effective exchange interaction, J$_{eff}^{Co-MnII}$, increases with $x$. This is because of the increasing hybridisations between nearest neighbour Co and MnII which couple parallely resulting in the weakened antiferromagnetic nearest neighbour MnI-MnII exchange interaction as the system gradually becomes MnI deficient (Co excess). The increasing ferromagnetic exchange interaction between nearest neighbour Ni and MnII and next nearest neighbour Ni-MnI and Ni-Co give rise to an increase in the overall ferromagnetic interaction. Upto $x=0.25$, there is a competition between the ferromagnetic and antiferromagnetic interactions which brings the T$_{c}$ down. Beyond $x=0.25$, the ferromagnetic components overwhelm the antiferromagnetic interactions resulting in the rise of the T$_{c}$ for higher values of $x$. In contrast, for (Mn$_{2-x}$Fe$_{x}$)NiGa, the J$_{eff}^{MnI-MnII}$ remains nearly constant with $x$. The strengths of the ferromagnetic Fe-MnII, Fe-Ni and MnI-Ni in these compounds are weaker in comparison to those in (Mn$_{2-x}$Co$_{x}$)NiGa with Ni-MnII exchange interaction having nearly same strengths and  larger than the Fe-MnII one, exact opposite to the Co-substituted compound. Such weaker ferromagnetic interactions are artefacts of weaker Fe-MnII hybridisations, particularly for low values of $x$ as can be seen from the atom-projected densities of states (Fig. 2, supplementary information) which shows that the major peaks of Fe and MnII are always separated. This weak interaction of MnII with one of the components in the 4a site keeps the strength of the interaction with the other component, MnI, at the same site intact across compositions although the concentration of MnI decreases gradually. This manifests itself in bringing down the T$_{c}$ considerably in (Mn$_{2-x}$Fe$_{x}$)NiGa for low $x$ and keeping it almost like that as $x$ increases. 

%\clearpage

For Mn$_{2}$(Ni$_{1-x}$Co$_{x}$)Ga and Mn$_{2}$(Ni$_{1-x}$Fe$_{x}$)Ga, we find the variations in the exchange interactions quite similar, qualitatively and quantitatively. The antiferromagnetic MnI-MnII interactions remain largely unaffected across compositions as the substitutions are not done in either of these sites. The inter-sublattice antiferromagnetic MnI-Fe and MnI-Co second-neighbour interactions are found to be substantial and identical in strengths. The strongest ferromagnetic interaction in Co-substituted system is that of Co-MnII pairs while the strength of Ni-MnII interaction is considerably weaker across the concentration range. In case of Fe-substituted system, though the strongest ferromagnetic interaction is due to Fe-MnII pairs for smaller $x$ values, the strength of Fe-MnII interactions quickly catch up with it. This is because of weakened hybridisations between Fe and MnII (Fig. 2, supplementary material) as $x$ increases. Thus, the initial decrease in T$_{c}$ for Mn$_{2}$(Ni$_{1-x}$Fe$_{x}$)Ga is due to weakening of the overall ferromagnetic interactions, primarily due to weak Fe-MnII hybridisations. For higher $x$, the J$_{eff}^{Ni-MnII}$ compensates for the J$_{eff}^{Fe-MnII}$, strengthening the ferromagnetic interactions in the system leading to an increase in T$_{c}$ with $x$. For Mn$_{2}$(Ni$_{1-x}$Co$_{x}$)Ga, the strong ferromagnetic exchange interactions for all $x$ values lead to an increase in T$_{c}$ with $x$. 

In systems with substitutions done at Ga sites, more number of interactions compete each other as Mn atoms are present at three different sites. The antiferromagnetic interactions are due to nearest neighbour MnI-MnII and MnI-MnIII and second neighbour MnII-MnIII. The ferromagnetic components in the exchange interactions are due to nearest neighbours X-MnIII (X=Co, Fe), X-MnII, Ni-MnII, Ni-MnIII and second neighbour Ni-MnI. For X=Co, that is when Ga is replaced with Co, the antiferromagnetic MnI-MnII interaction loses it's strength as does second neighbour antiferromagnetic MnII-MnIII interaction while the nearest neighbour MnIII-MnI interaction becomes more antiferromagnetic as concentration of Co increases. Among the ferromagnetic interactions, except Co-MnIII which decreases with concentration of Co, the other three increase. Strong hybridisations of Co with MnII and MnIII atoms are responsible for the strong ferromagnetic interactions in this system. This was predicted earlier \cite{MaPRB11} without explicit computations of the exchange interactions. Our calculation corroborates this with quantitative estimates. The antiferromagnetic exchange interactions behave the same way in systems with X=Fe. The ferromagnetic components for Fe-substituted system are much weaker than those for Co-substituted systems. In fact, although Fe-MnIII was the strongest ferromagnetic interaction initially, it drops fast giving way to Ni-MnII and Ni-MnIII.  Weaker hybridisations between Fe and Mn-II/Mn-III as Fe concentration increases can easily be seen from the atomic densities of states (Fig. 2 of Supplementary material). As Fe concentration increases, the Fe majority spin states move towards higher energies in comparison with MnII/MnIII states, thus making the hybridisations weaker. The weaker ferromagnetic components in Fe-substituted systems explain their smaller T$_{c}$ values, in comparison to Co-substituted systems, particularly for higher $x$ values. In both systems, the sharp decrease at $x=0.25$ is due to relatively stronger AFM interactions. As $x$ increases, the ferromagnetic interactions build up, effecting increase in T$_{c}$. 

\iffalse
%=====================================
\begin{table*}[t]
\centering
\caption{\label{table3} caption. }
\resizebox{\textwidth}{!}{%
\begin{tabular}{l c c c c c c c c@{\hspace{0.5cm}} c c c c c c c c c c c}
\hline \hline
 &\multicolumn{7}{c}{Cubic}
&&
\multicolumn{7}{c}{Tetragonal}
\\ \cline{2-8}\cline{10-16}
\vspace{-0.35 cm}
\\Systems  & M$_{\rm total}$ & M$_{\rm MnI}$ & M$_{\rm Fe/Co}$ & M$_{\rm MnII}$ & M$_{\rm Ni}$ & M$_{\rm MnIII}$& M$_{\rm Ga}$    && M$_{\rm total}$ & M$_{\rm MnI}$ & M$_{\rm Fe/Co}$ & M$_{\rm MnII}$ & M$_{\rm Ni}$ & M$_{\rm MnIII}$& M$_{\rm Ga}$  && $\Delta$M\\

\hline \hline
%\vspace{0.2 cm}
Mn$_{2}$NiGa  & 1.15 & -2.38 & - & 3.15 & 0.32 & - & 0.02    && 1.04 & -2.38 & - & 3.15 & 0.32 & - & 0.02  && 0.01\\

\hline \hline
\end{tabular}   
}
\end{table*}
\fi

\section{Summary and Conclusions}
With the help of first-principles density functional based calculations, we perform an in depth investigations of the effects of Fe and Co substitutions in magnetic shape memory system Mn$_{2}$NiGa. We study the site preferences of the substituents, the stabilities of the substituted compounds and their various properties in order to understand different aspects of substitutions in Mn$_{2}$NiGa, which in combination with available results on the Ni$_{2}$MnGa systems with similar substitutions can provide a consistent picture of the effects of such substitution in Ni-Mn-Ga alloys. We perform investigations mostly in the Hg$_{2}$CuTi structure which represents the high temperature austenite phase of Mn$_{2}$NiGa. Regarding site preferences and stability of the compounds formed by substitutions at different sites, we find that the substituents prefer the sites of substituting atoms when Ni or Mn is being substituted. In case of substitution of Ga, the substituents prefer to occupy the 4a sites in the Hg$_{2}$CuTi lattice, displacing the original constituent to the 4d site. This is in contrast with substituted Ni$_{2}$MnGa where site preferences sensitively depend upon the substituting site and the substituent.  We also find that the Co-substitution in Mn$_{2}$NiGa makes the system more stable in comparison to Fe-substituted Mn$_{2}$NiGa.

The patterns in the site occupancies lead to a gradual de-stabilisation of the martensitic phase of Mn$_{2}$NiGa irrespective of the site of substitution and the substituent, in agreement with experimental observations.\cite{MaPRB11, LuoJAP10} This uniform trend is an artefact of progressive weakening of the Jahn-Teller distortion that drives martensitic transformation in Mn$_{2}$NiGa. The Jahn-Teller instability in the Hg$_{2}$CuTi phase of Mn$_{2}$NiGa is due to the hybridisations between the $d$-states of constituents at the 4a and 4b positions(MnI and Ni respectively) and the $p$-states of Ga in the minority band. Since the substitution of another transition metal element like Fe or Co in Mn$_{2}$NiGa invariably replaces the elements at 4a and 4b positions, the hybridisations leading to Jahn-Teller instability, gradually vanishes with the increase in concentration of the substituent. The deep lying states of Fe and Co cannot restore the Jahn-Teller instability, rather contribute to it's decline. The same trend is observed in substituted Ni$_{2}$MnGa.\cite{ChunmeiPRB11} Thus, the site preferences of the substituents along with the positions of their states inside the minority spin bands are responsible for the martensitic transformation in Ni-Mn-Ga alloys over a large composition range.

The results on elastic modulii provide two useful information: (i) the tetragonal shear modulus C$^{\prime}$ can be considered as a predictor of the martensitic transformation and (ii) a correspondence between the ductile-to-brittle and metallic-to-covalent bonding transition can be curved out for substituted Mn$_{2}$NiGa. We find that the weakening of the martensitic stability largely correlates with the strengthening of the covalent bonds, due to hybridisations of the minority spin states of the substituents with either of the elements in 4a and 4b positions along with Ga at the 4d sites. The substitution at Ni sites render the systems more covalent as well as more brittle, while rest of the systems are, by and large, more metallic and ductile.

An immediate consequence of the disappearance of the Jahn-Teller distortion and the positions of the energy levels of the substituents which are deeper into the occupied parts of the minority spin bands of Mn$_{2}$NiGa, is opening of an energy gap in the minority band cutting through the Fermi level. This gap is like a half-metallic gap with spin polarisations of the substituted Mn$_{2}$NiGa reaching near $100\%$ when the substitution is complete. We find that all the compounds formed by $100\%$ substitution have nearly integer magnetic moment and nearly follow the Slater-Pauling rule of $M=N_{v}-24$, \cite{SlaterPRB36,PaulingPRB38} $M$ the total magnetic moment and $N_{v}$ the number of valence electrons. Thus Fe and Co substitutions, although leave Mn$_{2}$NiGa unsuitable for shape-memory applications except at low concentrations of the substituents, produce new compounds which are potentially useful for other magnetic applications.

The magnetic properties of Mn$_{2}$NiGa, in general improve, with more presence of the substituents. This is because of increasing ferromagnetic exchange interactions between the substituents and other magnetic atoms, and subsequent weakening of the dominant MnI-MnII antiferromagnetic interaction. The magnetic moments increase with concentration of the substituents as a result of this, the highest rise being in case of substitutions at the Ga sites, where an uncommon pattern of site occupancy magnifies ferromagnetic exchange interactions. Thus, we find a new stable magnetic material Mn$_{2}$NiCo with a moment as high as $\sim 9 \mu_{B} /f.u.$ The magnification of ferromagnetic exchange interactions elevate the Curie temperatures in these systems with T$_{c}$ of Mn$_{2}$NiCo as high as $\sim 900$ K. This hitherto unsynthesised material is in the league of newly discovered magnets in Heusler family having only $3d$ transition metals as their components.\cite{SanvitoSA17}

In conclusion, this work has explored the interrelations between the site occupancy, martensitic phase stability, bonding picture, mechanical and magnetic properties of Fe and Co substituted Mn$_{2}$NiGa. The results demonstrate that substituted Ni$_{2}$MnGa and Mn$_{2}$NiGa behave quite similarly and thus the effects of substitution of another transition metal in Ni-Mn-Ga system over a wide range of composition can be easily predicted. An important outcome of this work is the emergence of high spin-polarisable, nearly half-metallic compounds with high Curie temperatures upon complete substitution. A completely new material Mn$_{2}$NiCo shows promises with it's very high moment, spin polarisation and T$_{c}$. These results can motivate the experimentalists to explore such new materials. This work also acts as a guidance to the researchers about choice of suitable substituent and composition to improve functional properties of Ni-Mn-Ga systems.

\section*{ACKNOWLEDGMENT}
Authors would like to thank IIT Guwahati and DST India for the PARAM Supercomputing facility and the computer cluster in the Department of Physics, IIT Guwahati
\bibliographystyle{aip} % or  "apsrev4-1" "apsrmp4-1" "plain", "unsrt", %"alpha", "abbrv", etc.

\begin{thebibliography}{99}

\bibitem{UllakkoAPL96}
K.~Ullakko, J.~Huang, C.~Kantner, R.~O’handley, and V.~Kokorin,
\newblock Appl. Phys. Lett. {\bf 69}, 1966 (1996).

\bibitem{MurrayAPL00}
S.~J. Murray, M.~Marioni, S.~M. Allen, R.~C. O’Handley, and T.~A. Lograsso,
\newblock Appl. Phys. Lett. {\bf 77}, 886 (2000).

\bibitem{SozinovAPL02}
A.~Sozinov, A.~Likhachev, N.~Lanska, and K.~Ullakko,
\newblock Appl. Phys. Lett. {\bf 80}, 1746 (2002).

\bibitem{ChmielusNM09}
M.~Chmielus, X.~Zhang, C.~Witherspoon, D.~Dunand, and P.~M\"ullner,
\newblock Nature Mater. {\bf 8}, 863 (2009).

\bibitem{MarcosPRB02}
J.~Marcos, A.~Planes, L.~Ma\~nosa, F.~Casanova, X.~Batlle, A.~Labarta, and
  B.~Mart\'{\i}nez,
\newblock Phys. Rev. B {\bf 66}, 224413 (2002).

\bibitem{HuPRB01}
F.-x. Hu, B.-g. Shen, J.-r. Sun, and G.-h. Wu,
\newblock Phys. Rev. B {\bf 64}, 132412 (2001).

\bibitem{KrenkePRB07}
T.~Krenke et~al.,
\newblock Phys. Rev. B {\bf 75}, 104414 (2007).

\bibitem{PasqualePRB05}
M.~Pasquale, C.~P. Sasso, L.~H. Lewis, L.~Giudici, T.~Lograsso, and
  D.~Schlagel,
\newblock Phys. Rev. B {\bf 72}, 094435 (2005).

\bibitem{BiswasAPL05}
C.~Biswas, R.~Rawat, and S.~Barman,
\newblock Appl. Phys. Lett. {\bf 86}, 202508 (2005).

\bibitem{IngaleJAP09}
B.~Ingale, R.~Gopalan, V.~Chandrasekaran, and S.~Ram,
\newblock J. Appl. Phys. {\bf 105}, 023903 (2009).

\bibitem{AntoniJPCM09}
A.~Planes, L.~Mañosa, and M.~Acet,
\newblock J. Phys.: Condens. Matter {\bf 21}, 233201 (2009).

\bibitem{LikhachevMSE04}
A.~Likhachev, A.~Sozinov, and K.~Ullakko,
\newblock Materials Science and Engineering: A {\bf 378}, 513 (2004).

\bibitem{SozinovIEEE02}
A.~Sozinov, A.~Likhachev, and K.~Ullakko,
\newblock IEEE Trans. Magn. {\bf 38}, 2814 (2002).

\bibitem{WebsterPMB84}
P.~Webster, K.~Ziebeck, S.~Town, and M.~Peak,
\newblock Philos. Mag. B {\bf 49}, 295 (1984).

\bibitem{ParetiEPJB03}
L.~Pareti, M.~Solzi, F.~Albertini, and A.~Paoluzi,
\newblock The European Physical Journal B-Condensed Matter and Complex Systems
  {\bf 32}, 303 (2003).

\bibitem{SozinovAPL13}
A.~Sozinov, N.~Lanska, A.~Soroka, and W.~Zou,
\newblock Appl. Phys. Lett. {\bf 102}, 021902 (2013).

\bibitem{SotoPRB08}
D.~Soto et~al.,
\newblock Phys. Rev. B {\bf 77}, 184103 (2008).

\bibitem{KanomataPRB09}
T.~Kanomata et~al.,
\newblock Phys. Rev. B {\bf 80}, 214402 (2009).

\bibitem{KanomataIJAEM05}
T.~Kanomata, T.~Nozawa, D.~Kikuchi, H.~Nishihara, K.~Koyama, and K.~Watanabe,
\newblock International Journal of Applied Electromagnetics and Mechanics {\bf
  21}, 151 (2005).

\bibitem{GomesJAP06}
A.~Gomes, M.~Khan, S.~Stadler, N.~Ali, I.~Dubenko, A.~Takeuchi, and
  A.~Guimar{\~a}es,
\newblock J. Appl. Phys. {\bf 99}, 08Q106 (2006).

\bibitem{KhanJAP05}
M.~Khan, I.~Dubenko, S.~Stadler, and N.~Ali,
\newblock J. Appl. Phys. {\bf 97}, 10M304 (2005).

\bibitem{KanomataMetals13}
T.~Kanomata et~al.,
\newblock Metals {\bf 3}, 114 (2013).

\bibitem{SotoPM10}
D.~Soto-Parra et~al.,
\newblock Philosophical Magazine {\bf 90}, 2771 (2010).

\bibitem{KanomataPRB12}
T.~Kanomata et~al.,
\newblock Phys. Rev. B {\bf 85}, 134421 (2012).

\bibitem{LiuAPL05}
G.~Liu et~al.,
\newblock Appl. Phys. Lett. {\bf 87}, 262504 (2005).

\bibitem{LiuPRB06}
G.~D. Liu et~al.,
\newblock Phys. Rev. B {\bf 74}, 054435 (2006).

\bibitem{Kundumodulation17}
A.~Kundu, M.~E. Gruner, M.~Siewart, A.~Hucht, P.~Entel, and S.~Ghosh,
\newblock Phys. Rev. B. {\bf (in press)} (2017).

\bibitem{SinghAPL10}
S.~Singh, M.~Maniraj, S.~D’Souza, R.~Ranjan, and S.~Barman,
\newblock Appl. Phys. Lett. {\bf 96}, 081904 (2010).

\bibitem{BrownJPCM10}
P.~J. Brown, T.~Kanomata, K.~Neumann, K.~U. Neumann, B.~Ouladdiaf, A.~Sheikh,
  and K.~R.~A. Ziebeck,
\newblock J. Phys.: Condens. Matter {\bf 22}, 506001 (2010).

\bibitem{SinghAPL14}
S.~Singh, S.~Esakki~Muthu, A.~Senyshyn, P.~Rajput, E.~Suard, S.~Arumugam, and
  S.~Barman,
\newblock Appl. Phys. Lett. {\bf 104}, 051905 (2014).

\bibitem{LuoJAP10}
H.~Luo et~al.,
\newblock J. Appl. Phys. {\bf 107}, 013905 (2010).

\bibitem{MaPRB11}
L.~Ma et~al.,
\newblock Phys. Rev. B {\bf 84}, 224404 (2011).

\bibitem{DoCalphad08}
E.~C. Do, Y.-H. Shin, and B.-J. Lee,
\newblock Calphad {\bf 32}, 82 (2008).

\bibitem{RamamurthyJPCS86}
V.~Ramamurthy and S.~Rajendraprasad,
\newblock Journal of Physics and Chemistry of Solids {\bf 47}, 1109 (1986).

\bibitem{WinderJPCS58}
D.~Winder and C.~S. Smith,
\newblock Journal of Physics and Chemistry of Solids {\bf 4}, 128 (1958).

\bibitem{HuPRB09}
Q.-M. Hu, C.-M. Li, R.~Yang, S.~E. Kulkova, D.~I. Bazhanov, B.~Johansson, and
  L.~Vitos,
\newblock Phys. Rev. B {\bf 79}, 144112 (2009).

\bibitem{ChunmeiPRB10}
C.-M. Li, H.-B. Luo, Q.-M. Hu, R.~Yang, B.~Johansson, and L.~Vitos,
\newblock Phys. Rev. B {\bf 82}, 024201 (2010).

\bibitem{ChunmeiPRB11}
C.-M. Li, H.-B. Luo, Q.-M. Hu, R.~Yang, B.~Johansson, and L.~Vitos,
\newblock Phys. Rev. B {\bf 84}, 024206 (2011).

\bibitem{GhoshPBCM11}
S.~Ghosh, L.~Vitos, and B.~Sanyal,
\newblock Physica B: Condensed Matter {\bf 406}, 2240 (2011).

\bibitem{BarmanPRB08Ga2}
S.~R. Barman et~al.,
\newblock Phys. Rev. B {\bf 78}, 134406 (2008).

\bibitem{SanvitoSA17}
S.~Sanvito et~al.,
\newblock Science Advances {\bf 3}, e1602241 (2017).

\bibitem{PAW94}
P.~E. Bl\"ochl,
\newblock Phys. Rev. B {\bf 50}, 17953 (1994).

\bibitem{VASP196}
G.~Kresse and J.~Furthm\"uller,
\newblock Phys. Rev. B {\bf 54}, 11169 (1996).

\bibitem{VASP299}
G.~Kresse and D.~Joubert,
\newblock Phys. Rev. B {\bf 59}, 1758 (1999).

\bibitem{PBEGGA96}
J.~P. Perdew, K.~Burke, and M.~Ernzerhof,
\newblock Phys. Rev. Lett. {\bf 77}, 3865 (1996).

\bibitem{MP89}
M.~Methfessel and A.~T. Paxton,
\newblock Phys. Rev. B {\bf 40}, 3616 (1989).

\bibitem{EbertRPP11}
H.~Ebert, D.~Koedderitzsch, and J.~Minar,
\newblock Rep. Prog. Phys. {\bf 74}, 096501 (2011).

\bibitem{LiechtensteinJMMM87}
A.~Liechtenstein, M.~Katsnelson, V.~Antropov, and V.~Gubanov,
\newblock J. Magn. Magn. Mater. {\bf 67}, 65  (1987).

\bibitem{SokolovskiyPRB12}
V.~V. Sokolovskiy, V.~D. Buchelnikov, M.~A. Zagrebin, P.~Entel, S.~Sahoo, and
  M.~Ogura,
\newblock Phys. Rev. B {\bf 86}, 134418 (2012).

\bibitem{PaulJAP11}
S.~Paul and S.~Ghosh,
\newblock J. Appl. Phys. {\bf 110}, 063523 (2011).

\bibitem{BuchelnikovJPDAP15}
V.~Buchelnikov, V.~Sokolovskiy, M.~Zagrebin, M.~Tufatullina, and P.~Entel,
\newblock J. Phys. D: Appl. Phys. {\bf 48}, 164005 (2015).

\bibitem{SokolovskiyPRB15}
V.~V. Sokolovskiy, P.~Entel, V.~D. Buchelnikov, and M.~E. Gruner,
\newblock Phys. Rev. B {\bf 91}, 220409 (2015).

\bibitem{ChakrabartiPRB13}
A.~Chakrabarti, M.~Siewert, T.~Roy, K.~Mondal, A.~Banerjee, M.~E. Gruner, and
  P.~Entel,
\newblock Phys. Rev. B {\bf 88}, 174116 (2013).

\bibitem{LiuPRB08}
G.~Liu, X.~Dai, H.~Liu, J.~Chen, Y.~Li, G.~Xiao, and G.~Wu,
\newblock Phys. Rev. B {\bf 77}, 014424 (2008).

\bibitem{KandpalJPD07}
H.~Kandpal, G.~Fecher, and C.~Felser,
\newblock J. Phys. D: Appl. Phys. {\bf 40}, 1507 (2007).

\bibitem{FengPRB01}
Y.~Feng et~al.,
\newblock Phys. Rev. B {\bf 63}, 165109 (2001).

\bibitem{BurchPRL74}
T.~Burch, T.~Litrenta, and J.~Budnick,
\newblock Phys. Rev. Lett. {\bf 33}, 421 (1974).

\bibitem{HelmholdtJLCM87}
R.~Helmholdt and H.~Buschow,
\newblock J. Less-Common Met. {\bf 128}, 167 (1987).

\bibitem{PaulJPCM13}
S.~Paul, B.~Sanyal, and S.~Ghosh,
\newblock J. Phys.: Condens. Matter {\bf 25}, 236005 (2013).

\bibitem{PaulJAP14}
S.~Paul, A.~Kundu, B.~Sanyal, and S.~Ghosh,
\newblock J. Appl. Phys. {\bf 116}, 133903 (2014).

\bibitem{BarmanEPL07}
S.~R. Barman, S.~Banik, A.~K. Shukla, C.~Kamal, and A.~Chakrabarti,
\newblock EPL {\bf 80}, 57002 (2007).

\bibitem{WollmannPRB14}
L.~Wollmann, S.~Chadov, J.~K\"ubler, and C.~Felser,
\newblock Phys. Rev. B {\bf 90}, 214420 (2014).

\bibitem{AlijaniPRB11}
V.~Alijani, J.~Winterlik, G.~H. Fecher, S.~S. Naghavi, and C.~Felser,
\newblock Phys. Rev. B {\bf 83}, 184428 (2011).

\bibitem{LuoJAP08}
H.~Luo et~al.,
\newblock J. Appl. Phys. {\bf 103}, 083908 (2008).

\bibitem{Kundumn2fega17}
A.~Kundu and S.~Ghosh,
\newblock arXiv preprint arXiv: 1706.03425  (2017).

\bibitem{NiuSR12}
H.~Niu, X.~Chen, P.~Liu, W.~Xing, X.~Cheng, D.~Li, and Y.~Li,
\newblock Sci. Rep. {\bf 2}, 718 (2012).

\bibitem{PughPM54}
S.~Pugh,
\newblock Phil. Mag. {\bf 45}, 823 (1954).

\bibitem{VoigtAP89}
W.~Voigt,
\newblock Ann. Phys. (NY) {\bf 274}, 573 (1889).

\bibitem{PettiforMST92}
D.~Pettifor,
\newblock Mater. Sci. Tech. {\bf 8}, 345 (1992).

\bibitem{RoyPRB16}
T.~Roy, D.~Pandey, and A.~Chakrabarti,
\newblock Phys. Rev. B. {\bf 93}, 184102 (2016).

\bibitem{BarmanPRB08}
S.~Barman and A.~Chakrabarti,
\newblock Phys. Rev. B {\bf 77}, 176401 (2008).

\bibitem{MinakuchiJAC15}
K.~Minakuchi, R.~Umetsu, K.~Kobayashi, M.~Nagasako, and R.~Kainuma,
\newblock J. Alloys. Compds. {\bf 645}, 577  (2015).

\bibitem{SlaterPRB36}
J.~C. Slater,
\newblock Phys. Rev. {\bf 49}, 537 (1936).

\bibitem{PaulingPRB38}
L.~Pauling,
\newblock Phys. Rev. {\bf 54}, 899 (1938).

\end{thebibliography}

\end{document}